\documentclass[a4paper,11pt]{article}
\pdfoutput=1 

\usepackage{jheppub} 

\usepackage[T1]{fontenc} 
\usepackage{amsfonts}
\usepackage{tikz}
\usetikzlibrary{calc,math,patterns}
\usepackage{dcolumn}
\usepackage{bm}
\usepackage{physics}
\usepackage{subfig}
\newcommand{\nn}{\nonumber}
\newcommand{\bea}{\begin{eqnarray}}
	\newcommand{\ena}{\end{eqnarray}}
\newcommand{\inb}[3]{\left#1#2\right#3}

\makeatletter
\gdef\@fpheader{}
\makeatother

\def\cir{circle(3.5pt)}

\def\MyColorList{{"white","gray","black"}}
\tikzset{
	every picture/.style={thick},
	d/.style=
	{
		minimum width=2pt,
		inner sep=0pt,
		circle,
		fill=black
	},
	bline/.style={line width=0.6mm, gray},
	bnode/.style={fill=gray!50},
	eline/.style={dashed,thin},
	enode/.style={fill=pink!50,very thin}, 
rline/.style={line width=0.5mm, red},
shtriangle/.style={top color=gray!70},
squarednode/.style={rectangle, draw=red!60, fill=red!5, very thick, minimum size=5mm}
}

\tikzmath{
\r=1;
\dx=\r/2;
\h=sin(60)*\r;
\eps=0.3*\r;
}

\title{\boldmath $Z_3$ and
$(\times Z_3)^3$ symmetry protected topological paramagnets}

\author[a]{Hrant Topchyan,}
\author[b,c]{Vasilii Iugov,}
\author[a]{Mkhitar Mirumyan,}
\author[a]{Shahane Khachatryan,}
\author[a,d]{Tigran Hakobyan,}
\author[e,1]{Tigran Sedrakyan\note{Corresponding author.}}

\affiliation[a]{Alihkanyan National Laboratory, Yerevan Physics Institute, Alikhanian Br. 2, 0036, Yerevan, Armenia}
\affiliation[b]{Simons Center for Geometry and Physics, Stony Brook University, Stony Brook, NY, 11794-3636, USA}
\affiliation[c]{C. N. Yang Institute for Theoretical Physics, Stony Brook University, Stony Brook, NY, 11794-3636, USA}
\affiliation[d]{Yerevan State University, Alex Manoogian 1, 0025, Yerevan, Armenia}
\affiliation[e]{Department of Physics, University of Massachusetts, Amherst, Massachusetts 01003, USA}
\emailAdd{hranttopchyan1@gmail.com}
\emailAdd{vasyayugov@gmail.com}
\emailAdd{mkhitar@gmail.com}
\emailAdd{shah@yerphi.am}
\emailAdd{tigran.hakobyan@ysu.am}
\emailAdd{tsedrakyan@umass.edu}

\abstract{We identify two-dimensional three-state Potts paramagnets with gapless edge modes on a triangular lattice protected
by $(\times Z_3)^3\equiv Z_3\times Z_3\times Z_3$ symmetry and smaller  $Z_3$ symmetry. We derive microscopic models for the gapless edge, 
uncover their symmetries and analyze the conformal properties.
We study the properties of the gapless edge by employing the numerical density-matrix renormalization group (DMRG) simulation and exact diagonalization. We discuss the corresponding conformal field theory, its central charge, and the scaling dimension of the corresponding primary field. We argue, that the low energy limit of our edge modes
defined by the $SU_k(3)/SU_k(2)$ coset conformal field theory with the level $k=2$.
The discussed two-dimensional models 
realize a variety of symmetry-protected topological phases, opening a window for studies of the unconventional quantum criticalities between them.}

\begin{document} 
\maketitle
\flushbottom

\section{Introduction}
\label{sec:intro}
Over the past decade, symmetry-protected topological (SPT)
phases \cite{Gu-Wen-2009, Senthil,Wen-2011b,Wen-2011c,Wen-2011d,Berg-2009,Berg-2010} have generated a lot of research interest
\cite{Wen-2012,Levin-Gu,Wen-2014,Gu-Wen-2014,Wen-2013,Wu-2013,Yoshida-2015,Yoshida-2016-1,Yoshida-2017,Furusaki-2014,Wang-2014,Wang-2015,DHLee-2017,Tantivasadakarn-2017,Hsieh-2020,Wei-2018,Motrunich-2014,Wang-2016,Wang-2018,Fendley-2020,Kapustin1,Kapustin2,Gaiotto,Cheng-2020,Alavirad-2021,Lanzetta-2022}. They are essentially different from Landau definition of phases via local order parameters and carry topological characterization.

It is well-established that symmetry-broken ordered states are all characterized by group theory describing the order parameter manifold. Such states are short-range entangled. The SPT states are also short-range entangled\cite{Vishwanath}, in contrast to topologically ordered states \cite{TO1,TO2,TO3,TO4,TO5,TO6,TO7,TO8,TO9,TO10,TO11, TSChenan, nature, Sedrakyan-2012} that are long-range entangled. Thus, short-range entangled phases are generally symmetry broken (described by the Landau paradigm), SPT (outside of the Landau paradigm), or support simultaneous coexistence of symmetry-breaking and SPT order. Importantly, SPT orders support the symmetry-protected gapless boundary excitations. These states, often with non-standard statistics, are substantial for the basics of topological quantum computation.
The topological systems near criticality are generally remarkable for their universal finite-size scaling behavior \cite{Kamenev-2016,Kamenev-2020} and sensitivity to symmetry-breaking perturbations
\cite{Sedrakyan-2020,Sedrakyan-2022-1}. They may possess both paramagnetic  \cite{Levin-Gu}  and spin-ordered phases, with the latter being closely related to Neel orders
\cite{Sedrakyan-2017,Sedrakyan-2018,Sedrakyan-2022-2}.

The topological insulators of free fermions, which feature an SPT phase protected by $U(1)$ and time-reversal symmetry, are well-studied in the literature.
There are two distinct types of such time-reversal invariant band insulators: topological insulators and conventional insulators \cite{Kane-2005-1,Kane-2005-2,Balents-2007,Hasan-2010}. The two families of insulators are distinguished by the fact that topological insulators have
protected gapless boundary modes, while trivial insulators do
not. This is because time reversal and charge conservation symmetry play a crucial role in this physics: the boundary modes will gain a gap if any of these two symmetries are broken explicitly or spontaneously. Then the distinction between topological insulators and conventional insulators disappears.

A deeper understanding of SPT phases followed the work of
Xiao-Gang Wen and collaborators, where a classification of SPT states was formulated based on cohomology classes of discrete groups \cite{Gu-Wen-2009,Wen-2011b,Wen-2011c,Wen-2011d,Wen-2013, Gu-Wen-2014}. 
Full classification of SPT phases in 1D system was presented in  \cite{Wen-2011b,Wen-2011c,Cirac-2011,Kitaev-2011}
This classification has become a powerful mathematical tool for characterizing and distinguishing between different SPT states, and predicting their properties, such as the number of protected edge modes and the nature of the ground state degeneracy. 
Along these lines of research, an interesting and important result was reported in Ref.~\cite{Levin-Gu}, where the authors show how an ordinary paramagnetic Ising model with $Z_2$ symmetry can be modified to produce a gapped SPT state with gapless $Z_2$ symmetry-protected edge states.

According to the standard definition, gapped SPT states have several characteristic properties. The system should have an internal symmetry $G$ and a ground state with no spontaneous symmetry breaking. The SPT state should be distinct from the "trivial" state (a kind of product spin/boson state) and can not be continuously (in some parameters) connected with a "trivial" state without closing the gap in the bulk. The last property is that the two states can be connected continuously without closing the gap, but one or more symmetries of the Hamiltonian should be broken. As a general rule, an SPT system has massless edge states, which at some momentum have zero energy. Furthermore, the presence of the gapless edge (or its absence) can, in principle, distinguish the SPT state from the trivial ground state. Therefore, some characteristic classes topologically protect these edge states. Refs.~\cite{Wen-2012,Wen-2013,Wen-2014} report classification of SPT states in $d$ dimensional spin models according to
cohomology groups $H^{d+1}(G, U(1))$ of the site symmetry group $G$
of the models with coefficients in wave functions phase factor $U(1)$. 
This bulk-boundary link and appearance of massless edge states is a hallmark of t'Hooft anomaly \cite{t'Hooft}, linked to well known Lieb-Schultz-Mattis (LSM) theorem,  which is an anomaly between translational invariance and global symmetry G of the system \cite{Alavirad-2021,Lanzetta-2022}. In Ref.~ \cite{Lanzetta-2022}, it was shown how the presence of this anomaly helps to formulate the 
low energy limit of lattice theories. Namely, if the anomaly symmetry of the lattice model coincides with the one in quantum field theory, then both low energy limits may also coincide.

One approach to understanding SPT phases is through the use of group cohomology, that allows both for the characterization of topological phases in terms of the underlying symmetry group and their construction\cite{Wen-2013}.  Specifically, Ref. \cite{Wen-2013} examines how the concept of a group extension can be used to construct SPT models with a variety of different symmetries, including discrete and continuous symmetries. Here one has to derive an explicit group cocycle and then construct a lattice SPT model following the procedure outlined in that work. 
The group cohomology classification also offers physical insites into the implications of these models, including their relevance to topological insulators, superconductors, and other exotic quantum materials. 

In this respect, Ref.~\cite{Levin-Gu} also reports a nice procedure formulated to generate massless edge modes in the paramagnetic phase of two dimensional $Z_2$ Ising model in bulk. One can make any unitary transformation of the Hamiltonian and operators and generate a new theory in bulk. Since the transformation is unitary,
the spectrum in the bulk of the model will remain the same. However, the excitations may gain nontrivial statistics upon gauging\cite{Wu-2013,Yoshida-2015,Yoshida-2016-1,Yoshida-2017} .
 Within this approach, to find the Hamiltonian of the boundary modes, one fixes the operators of unitary transformation on the system boundary by fixing external spins and summing up those spins for a new Hamiltonian to have the same symmetries as the parent one. Due to this summation, the new Hamiltonian is not unitary equivalent to the initial one but differs only by the appearance of new edge states, which can become gapless. Applying this type of transformation to the paramagnetic $Z_2$ Ising model, Levin and Gu have identified the edge states with the one-dimensional (1D) Hamiltonian of the gapless
$XX $ model (conformal theory with central charge $c=1$). Hence, this transformed 2D model with a gapless edge forms an SPT paramagnet.

In this paper, we construct an SPT phase on a two-dimensional triangular lattice based on the higher spin system, namely the three-state Potts model in its paramagnetic phase. We discuss the construction of the model in detail and formulate the SPT lattice paramagnets with $Z_3$ symmetry. We start from the three-state Potts model on a triangular lattice in the paramagnetic phase, whose spectrum is gapped. The paramagnetic Potts model has larger, $Z_3 \times Z_3 \times Z_3$ symmetry, where each $Z_3$ is defined on three different triangular lattice sites. The triangular lattice contains three triangular sub-lattices of larger size, denoting with index $i=1,2,3$. Then, following the approach of Ref.~\cite{Levin-Gu}, we reformulate the $Z_3$ Potts model using symmetry-protected unitary transformation, which has nontrivial cocycle property in the $Z_3 \times Z_3 \times Z_3$ and $Z_3$ groups. According to Refs.~ \cite{Wen-2012, Wen-2013, Wen-2014}, the variety of SPT states in 2D models is defined by elements of the third cohomology group
of the symmetry group of the model. In our case it leads to $H^3(Z_3 \times Z_3 \times Z_3, U(1)) = (\times Z_3)^7$ \cite{Wu-2013,Tantivasadakarn-2017} and $H^3(Z_3, U(1)) = Z_3$. We show how
$ Z_3 \times Z_3 \times Z_3  $ and $Z_3$ symmetric unitary transformations
preserve the spectrum in bulk and lead to the appearance of gapless edge states manifesting the presence of the SPT phase. Models  with  $Z_3$
and $ Z_3 \times Z_3 \times Z_3  $ symmetry were discussed also in \cite{Wang-2015}  and \cite{Alavirad-2021, Wu-2013} respectively.
Finally, we identify the Hamiltonian operators describing the edge states in $Z_3 \times Z_3 \times Z_3$ symmetric case and investigate their symmetry and conformal properties.

Importantly, we found that our edge Hamiltonian has hidden $U(1)$ anomalous symmetry, which is responsible for the current algebra of the corresponding low-energy CFT and shrinks the possible theories considerably, helping us to identify the low-energy CFT candidate as the coset $SU(3)_2/SU(2)_2$ model.
We have used the technique developed in \cite{Wang-Zou-2022} for $U(1)$ subgroup in XXZ Heisenberg chain to detect corresponding Kac-Moody algebra and its central extension in our model.

\section{A Three-state Potts model}
The Hamiltonian of the three-state Potts model can be derived from its action formulation following the standard prescription \cite{Kogut-1979}
\begin{align}
	\label{Z3Potts}
	&H_P=\gamma \sum_{{\bf r}\in 2D,{\bf \mu}} \big(\varepsilon^{S^z_{\bf r}}\varepsilon^{-S^z_{\bf r+\bf \mu^a}}
	+\varepsilon^{-S^z_{\bf r}}\varepsilon^{S^z_{\bf r+\bf \mu}} \big)
	- \sum_{{\bf r}\in 2D} (X^+_{\bf  r}+X^-_{\bf r}),
	\\
	\label{XS}
	&X^+_{\bf r}=(X^-_{\bf r})^\dagger =
	\left(
	\begin{array}{ccc}
		0&1&0\\
		0&0&1\\
		1&0&0
	\end{array}
	\right),
	\qquad
	S^z_{\bf r}=
	\left(
	\begin{array}{ccc}
		1&0&0\\
		0&0&0\\
		0&0&-1
	\end{array}
	\right),
\end{align}
where $\varepsilon=e^{2\pi i/3}$ is a basic element of the group $Z_3$,
${\bf \mu^a},\; a=1,2,3$ are three basic lattice vectors of the triangular lattice, and $\gamma$ is the interaction constant. The
operators $X^{\pm}_{\bf r}$ and $\varepsilon^{\pm S^{z}_{\bf r}}$ belong to the cyclic spin-1 representation of quantum group $SU_q(2)$ with $q=3$ and they obey the algebra
\bea
\label{QGalgebra}
X^{\pm}_{\bf r} \varepsilon^{S^z_{\bf r}}=\varepsilon^{\mp 1}  \varepsilon^{S^z_{\bf r}} X^{\pm}_{\bf r},\quad (X^+_{\bf r})^3=(X^-_{\bf r})^3=1.
\ena
Note that the above matrices are related to the parafermion generators\cite{chern}.

We consider the three-state Potts model on a triangular lattice, $R$, with boundary, $\partial R$, in the paramagnetic phase when $\gamma=0$. Then the Hamiltonian is simple and divided into two non-interacting parts on the boundary and in bulk:  $H_P=H_{bulk}+H_{edge}$
\bea
\label{ham}
H_{bulk}=- \sum_{{\bf r}\in R} (X^+_{\bf r}+X^-_{\bf r}), \quad H_{edge}=- \sum_{{\bf r}\in \partial R} (X^+_{\bf r}+X^-_{\bf r}).
\ena
The Hamiltonian (\ref{ham}) has a much larger symmetry in its paramagnetic phase. The sites of the triangular lattice can be divided
into three groups. Each site of the basic triangles can be included as a site in one of three larger triangular lattices. Therefore the $Z_3$ Ising
symmetry group can be written as
\bea
\label{Ising-1}
X^\pm=\prod_{{\bf m}, i=1,2,3} X^\pm_{{\bf m},i},
\ena
where ${\bf m}$ denotes the small triangles of the original triangular lattice, while $i=1,2,3$ correspond to the three triangular sub-lattices.
It can also be seen that the paramagnetic Potts model (\ref{ham})
has larger symmetry based on the generators
\bea
\label{Ising-3}
X^\pm_i=\prod_{{\bf m}} X^\pm_{{\bf m},i} \quad\text{with}\quad  i=1,2,3,
\ena
defining the $Z_3 \times Z_3 \times Z_3$ symmetry.
Below we will discuss the possibility of having protecting symmetries in the paramagnetic Potts model (\ref{ham}) and the related construction of SPT states in detail.

\section{Classification of SPT states via cohomologies}

The SPT states in two dimensions can be classified via the third $U(1)$-valued cohomology group of the protecting symmetry group of the model~\cite{Wen-2013,Gu-Wen-2014,Kapustin1,Kapustin2,Gaiotto}. Below, we will discuss the basics of the cohomologies of finite abelian groups and apply them to $Z_3$ and $Z_3 \times Z_3 \times Z_3$ symmetric paramagnets.

\subsection{Introduction to cohomologies of finite abelian groups and application to $Z_3$ and $Z_3 \times Z_3 \times Z_3$ groups}

Here we briefly describe the group cohomologies. We
confine ourselves to finite abelian groups considered in the present work.

Let $n$ represent the elements of a finite abelian group $G$ in its additive form. For the group $Z_3$, its elements, $n$, coincide with $S^z$  defined above for the Potts model. The $U(1)$ phase-valued function $\omega=e^{i\psi}$ on $k$ group
variables is called a $k$-dimensional cochain. The space of all $k$-cochains, denoted by $C^k$,
form an abelian group inherited from the $U(1)$. Note that for arbitrary $G$, cochains form not a vector space but a mere abelian group.
Nevertheless, in this paper, we are working with $G=(\times Z_3)^n$, which is a vector space over the field $Z_3$, and we will sometimes use vector space language for it.

The following coboundary operator maps the $C^k$ to $C^{k+1}$ preserving the group structure
(here, the linearity)
\cite{Wen-2012,Wen-2013,Gu-Wen-2014, Feigin-1989}:
\begin{align}
	\label{dpsi}
	\delta\psi(n_1,\dots,n_{k+1})&=\psi(n_2,\dots,n_{k+1})
	-\psi(n_1+n_2,n_3,\dots, n_{k+1})
	+\psi(n_1,n_2+n_3,\dots, n_{k+1})
	\nonumber
	\\
	&+\ldots +(-1)^k \psi(n_1,\dots,n_{k-1},n_k+n_{k+1})
	+(-1)^{k+1}\psi(n_1,\dots,n_k).
\end{align}
This  operator is nilpotent as one can verify that
\begin{equation}
	\label{d^2}
	\delta^2 \psi=0 \mod 2\pi, \quad \text{or} \quad \delta^2 \omega=1.
\end{equation}
Its form is inherited from the more conventional action on the $G$-invariant
cochain  $\nu=e^{i\varphi}$, containing one more argument:
\begin{align}
	\label{nu}
	\varphi(n_0,\dots,n_k)&=\psi(n_1-n_0,\dots, n_k-n_{k-1}),
	\\
	\label{dnu}
	\delta \varphi(n_0,\dots,n_{k+1})&=\sum_{l=0}^{k+1}(-1)^l \varphi(n_0,\dots,\hat{n}_{l},\dots,n_{k+1}),
\end{align}
where the "hat" marks the absence of the underlying argument.
The  coboundary of the lowest cochains is listed above in the text \eqref{cochain-2}, \eqref{cochain-3},
and \eqref{cochain-4}.

In the  lowest dimensions,
the coboundary operator \eqref{dpsi}
takes the following form in the $\omega$ notations,
\bea
\label{cochain-2}
\delta \omega_1(n_1,n_2)&=&\frac{\omega_1(n_2) \omega_1(n_1)}{\omega_1(n_1+n_2)},
\\
\label{cochain-3}	
\delta \omega_2 (n_1,n_2,n_3)&=&\frac{\omega_2(n_2,n_3) \omega_2(n_1,
	n_2+n_3)}{\omega_2(n_1+n_2, n_3)\omega_2(n_1, n_2)},
\\
\label{cochain-4}
\delta \omega_3(n_1,n_2,n_3,n_4)&=&\frac{\omega_3(n_2,n_3,n_4) \omega_3(n_1, n_2+n_3,n_4) \omega_3(n_1,n_2,n_3)}{\omega_3(n_1+n_2,n_3,n_4)\omega_3(n_1,n_2,n_3+n_4)}.
\ena

\begin{figure}[t]	
	\centering
	\includegraphics[width=110mm]{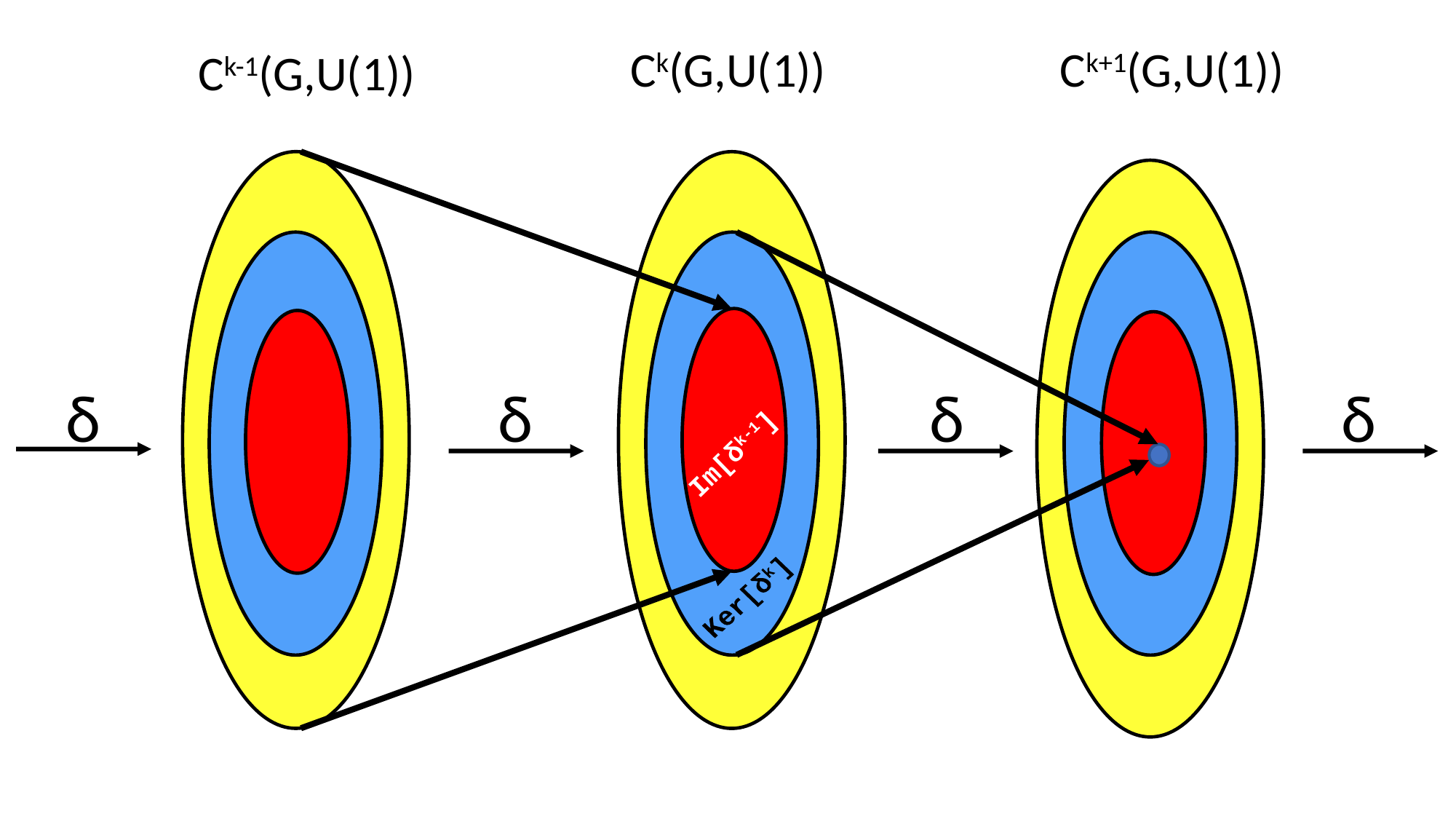}
	\caption{\label{fig-seq} The sequence of coboundary maps $\delta^k$ acting on the cochain spaces $C^k$
		depicted by large disks. The medium disks are cocycles, and the small, red disks are coboundaries.
		The cohomologies $H^k$ are represented by blue areas.
	}
\end{figure}

A cochain with trivial  coboundary is called a cocycle,
\bea
\label{cocycle}
\delta \omega=1, \quad \text{or} \quad \delta \psi=0\mod 2\pi.
\ena
In particular, a one-cocycle, $ \omega_1$, forms a one-dimensional representation of the group
as follows from \eqref{cochain-2} and \eqref{cocycle}.
At the same time, according to \eqref{cochain-3}, a two-cocyle describes a projective representation $t_n$:
$$
t_{n_1}t_{n_2}=\omega(n_1,n_2)t_{n_1+n_2}.
$$
Then the cocycle condition $\delta \omega_2=1$ implies the associativity of the product $t_{n_1}t_{n_2}t_{n_3}$.

A cochain $\omega_k=\delta \omega_{k-1}$ is called a coboundary.
Using \eqref{d^2}, one can see that the coboundaries obey the cocycle condition, but the converse statement is generally not true.
In other words, certain cocycles may exist which are not coboundaries.
Alike the cochains, the cocycles and coboundaries form abelian groups.
Note that the property \eqref{d^2} highlights the similarity between cochains and differential forms.
The operator $\delta$ is in direct analogy to the differentiation operator $\dd$ on
differential forms on a smooth manifold. In this analogy, cocycles and cochains correspond to closed and exact forms.
We will sometimes use this language later in the text.

Similarly to de-Rham cohomologies, the cohomology group is the factor space of the cocycles
over exact cochains. More precisely,
denoting the coboundary operator acting on the $k$-cochains by $\delta^k$,
the $k$-th cohomology group is given by the factor group
\bea
\label{Hq}
H^k(G,U(1))=\frac{\text{Ker}\,\delta^k}{\text{Im}\,\delta^{k-1}},
\ena
where we use the standard notations for the kernel and image of the map.
The sequence of the $k$-cohomologies with the coboundary operator action is depicted in Fig.~\ref{fig-seq}.

Below, we will apply the approach to our case of $Z_3$ and $Z_3\times Z_3\times Z_3$  groups.
According to the statement formulated in Refs.~\cite{Wen-2013,Gu-Wen-2014},  the SPT states in two dimensions
are classified by the third $U(1)$-valued cohomology group of the symmetry group of the model.
For the cyclic groups $Z_3$ and its threefold  product, the corresponding cohomologies are given by
\begin{align}
	\label{H3Z3}
	H^3(Z_3, U(1))&=Z_3,
	\\
	H^3(Z_3 \times Z_3 \times Z_3, U(1))&= (\times Z_3)^7.
	\label{H3Z3^3}
\end{align}
The nontrivial three-cocycles in both groups were presented in papers
\cite{Wen-2013,Yoshida-2015,Wu-2013,Tantivasadakarn-2017,Coste-2000,Moore-1989,Wakui-1992}.
A mathematically straightforward way
of finding the basic elements of $H^3$ paves the way to formulating the appropriate unitary transformation of paramagnetic
$Z_3$ Potts model and finding the Hamiltonian of massless edge modes there.


\subsection{Third cohomology and cocycles of $Z_3$ group}
In this subsection, we consider in detail the structure of the third cohomology of the cyclic group $Z_3$
characterized by the same group
\eqref{H3Z3}. Although the structure is known in the mathematical literature, one can clarify it from the following
observation without a detailed reference to the group cohomology theory.

Let us look for the nontrivial cocycle, which generates the cohomology group $H^3$ in the form
$\omega=\varepsilon^{i\psi}$, where  $\psi$ is a polynomial in qutrit variables $n_1,n_2,n_3$ with integral modulo $3$ coefficients.
In particular, according to \eqref{cochain-2}, the cocycles  in $H^1$  obey $\omega_1(n)=\omega_1(n+3)=\omega_1(n)\omega_1(3)$,
so that $\omega(3)=1=\omega(0)$. The periodicity of a two-cocycle imposes that the function $\omega(n_1,3)=\omega(n_1,0)$
is a constant and coincides with unity. The same is true for the second argument.
So, any $Z_3$ $k$-cocycle may be expressed as $\omega_k=\varepsilon^{\psi_k}$,
where $\psi_k$ is a polynomial in $n_1,\dots,n_k$ with modulo $3$ coefficients.
In contrast, the periodicity is not obligatory for the cochains, which generally
take values in the $U(1)$ group.
In particular, a cochain for the coboundary
$\omega_2=\delta\omega_1=\varepsilon^{n_1n_2(n_1+n_2)}$ can be given by
$\omega_1(n)=\varepsilon^{\frac13 n^3}$, which one can 
check.  We see that the function $\omega_1$ has a  higher period equal to $9$.
The given example agrees with the triviality of the second cohomology of $Z_3$.
In fact, the $H^3$ is generated by the cocycle
\bea
\label{cocycle-Z3}
\omega_3=\varepsilon^\psi \quad \text{with} \quad \psi(n_1,n_2,n_3)=n_1n_2n_3(n_2+n_3).
\ena
The cocycle
condition $\delta\psi=0$ modulo 3 is possible to verify using \eqref{dpsi} or \eqref{cochain-4}.
From the other side, $\psi$ is not a coboundary, i.e., there is no a 2-cochain which obeys $\delta\omega_2=\varepsilon^\psi$.
Indeed, taking the two different values for the arguments as $n_1=n_2=n_3=\pm1$ and using the equation \eqref{3-cocycle-3},
we get the following two contradictious conditions:
$\omega_2(1,-1)/\omega_2(-1,1)=\varepsilon^{\pm1}$.
Note that the cocycle $\psi'$, obtained from $\psi$ by permutation of $n_1$ and  $n_2$, belongs to the same
cohomology class since $\psi'-\psi=\delta \alpha$
with the 2-cochain $\alpha=(n_1+n_2)^2n_1n_2$.

Next,  the three-cocycle $\omega=\varepsilon^{n_1n_2n_3}$ is trivial since is the
coboundary of the 2-cochain $\psi=n_1^2n_2$.
Indeed,
\bea
\label{LG-Z3}
\delta \varepsilon^\psi=\varepsilon^{n_2^2n_3-(n_1+n_2)^2n_3+n_1^2(n_2+n_3)-n_1^2n_2}=\varepsilon^{-2n_1n_2n_3}=\varepsilon^{n_1n_2n_3}.
\ena
Note that in $Z_2$ case when $\varepsilon=-1$, and $n_i=0,1$ label qubit values, this cocycle is nontrivial
since, as is possible to verify,  $\delta \varepsilon^\psi=1$.
The product of two cocycles \eqref{LG-Z3} and \eqref{cocycle-Z3}
gives rise to the conventional generator of the third cohomology group,
\bea
\label{theta}
\omega_3'=\varepsilon^{- n_1n_2n_3(1+n_2+n_3)}=\varepsilon^{n_1\left[\frac{n_2+n_3}{3}\right]}.
\ena
Here $[x]$ is an integral part of $x$, and one has to set the values $n_i=0,1,2$ firmly.
Indeed, one can verify by direct substitution that both expressions are equivalent since
$$
n_2n_3(1+n_2+n_3)=2\left[\frac{n_2+n_3}{3}\right] \mod 3.
$$
The translation invariance is
an advantage of the first expression for the cocycle we introduced.
In contrast, the standard form on the right is restricted to the
qutrit values $n_i=0,1,2$ only.
Note that unity  in the exponent can be omitted since it produces a coboundary multiplier
as in \eqref{LG-Z3} so that an equivalent representative may be used instead,
$\omega\sim\varepsilon^{- n_1n_2n_3(n_2+n_3)}$.

\subsection{Third cohomology and cocycles for $Z_3 \times Z_3 \times Z_3$ group}
In this subsection, we analyze the symmetry given by the threefold product of the $Z_3$ group and discuss the corresponding
cohomology and cocycles, discussing these within the context of the SPT phases.
It is convenient to label the group elements by a vector  $n=(n^{(1)},n^{(2)},n^{(3)})$, each component
of which labels the corresponding $Z_3$ term in the group direct product.
We will refer to the upper index in $n^{(j)}$ as to the {\em color} of this variable.
Any cochain can be written in the form $\omega_2(n_1,n_2)=\varepsilon^{\psi_2(n_1,n_2)}$ with $\psi_2$ being some polynomial in $n_{1,2}^{(j)}$. Since in $Z_3$, the cubic and higher orders monomials are reducing to the first and second-order,
$\psi_2(n_1,n_2)$ is a polynomial of degree not greater than two in each of the variables $n_{1,2}^{(j)}$.

We want to find all of the nontrivial 3-cocycles, i.e., closed 3-cochains that are not exact. One of them is
\bea
\label{3-cocycle-1}
\omega^{123}_3(n_1,n_2,n_3)=\varepsilon^{n^{(1)}_1 n^{(2)}_2 n^{(3)}_3}.
\ena
The cocycle condition, $\delta\omega^{123}_3=1$, is possible to check using the definition \eqref{cochain-4}.
It cannot be expressed as a linear combination of 3-coboundaries; thus, it is nontrivial.
For a more detailed explanation, see Appendix \ref{sec:nontriviality}.

To check that a cochain is nontrivial, we use machine computations using Mathematica software, as there are too many 3-coboundaries to write them down.
But, at least for the cocycle \eqref{3-cocycle-1}, there is a simple way to show that it is nontrivial. Consider any trivial 3-cocycle $\omega_3=\delta \omega_2$.
Then a simple calculation shows that
\begin{equation}
	\frac{\omega_3(n_1,n_2,n_3)\omega_3(n_2,n_3,n_1)\omega_3(n_3,n_1,n_2)}{\omega_3(n_2,n_1,n_3)\omega_3(n_3,n_2,n_1)\omega_3(n_1,n_3,n_2)}=1.
\end{equation}
However, for $\omega_3^{ijk}$ that is not the case:
\begin{equation}
	\left.\frac{\omega_3^{ijk}(n_1,n_2,n_3)\omega_3^{ijk}(n_2,n_3,n_1)\omega_3^{ijk}(n_3,n_1,n_2)}{\omega_3^{ijk}(n_2,n_1,n_3)
		\omega_3^{ijk}(n_3,n_2,n_1)\omega_3^{ijk}(n_1,n_3,n_2)}\right|_{\substack{n_1=(1,0,0)\\n_2=(0,1,0)\\n_{3}=(0,0,1)}}=\varepsilon\ne 1.
\end{equation}
Thus, $\omega_3^{ijk}$ is not a coboundary.
Unfortunately, the remaining six nontrivial 3-cocycles cannot be shown to be nontrivial as straightforwardly. Computations show that
\bea
\label{3-cocycle-2}
\omega^{i}_3(n_1,n_2,n_3) = \varepsilon^{n^{(i)}_1 n^{(i)}_2 n^{(i)}_3 \big( n^{(i)}_2+ n^{(i)}_3\big)},
\qquad i=1,2,3
\ena
are not coboundaries. However, using Eq.~(\ref{cochain-4}), one can check that its boundary is
$\delta\omega^{i}_3(n_1,n_2,n_3,n_4)=1$. Hence, it is a nontrivial cocycle. There are 3 different cocycles of this type.

Another three nontrivial cocycles can be written in the form
\bea
\label{3-cocycle-3}
\omega^{ij}_3(n_1,n_2,n_3)
=\varepsilon^{n^{(i)}_1n^{(j)}_2 n^{(j)}_3\big( n^{(j)}_2 +  n^{(j)}_3\big)}.
\ena
They are not exact, and by use of formula (\ref{cochain-4}) for coboundary, one can check
that $\delta \omega_3^{ij}=1$. Hence, considering the different pairs of colors $(i,j)$ with $i<j$,
we get another three nontrivial cocycles.
As per the above discussion, altogether we have $1+3+3=7$ nontrivial cocycles defined by Eqs.~(\ref{3-cocycle-1}, \ref{3-cocycle-2}, \ref{3-cocycle-3}).
This is in complete agreement with the Eq.~\eqref{H3Z3^3}.
It is left to check that those cocycles constitute a system of linearly independent elements of $H^3$. Note that we can split the vector spaces $C^k$ of $k$-cocycles into subspaces formed by polynomials with monomials consisting of given colors:
\begin{equation}
	\label{forms_splitting}
	C^k = C^k_{(123)} \times C^k_{(12)} \times C^k_{(13)} \times C^k_{(23)} \times C^k_{(1)}\times C^k_{(2)}\times C^k_{(3)} \times C^k_{()}
\end{equation}
For example, $C^k_{(ij)}$ is spanned by monomials each of which contains both $n^{(i)}_k$ and $n^{(j)}_l$ for some $k$ and $l$. Forms in $C^k_{()}$ must be constant. Thus, $C^k_{()}$ is trivial and is spanned by a constant form. The coboundary operator $\delta: C^k \to C^{k+1}$ splits into operators acting between corresponding subgroups of $C_k$:
\begin{equation}
	\begin{split}
		\delta C^k_{(123)} \subset C^{k+1}_{(123)}, \quad
		\delta C^k_{(ij)}  \subset C^{k+1}_{(ij)}, \quad
		\delta C^k_{(i)}  \subset C^{k+1}_{(i)}, \quad
		\delta C^k_{()} \subset C^{k+1}_{()}
	\end{split}
\end{equation}
Thus, we can split the $k$-dimensional cohomology groups into a direct product of groups
\begin{equation}
	H^k(G,U(1)) = H^k_{(123)}\times  H^k_{(12)} \times  H^k_{(13)}\times  H^k_{(23)} \times  H^k_{(1)}\times  H^k_{(2)} \times  H^k_{(3)} \times  H^k_{()}
\end{equation}
The 3-cocycles we found earlier belong to different parts of that product. Thus, they are an independent set and represent a full basis in $H^3(Z_3\times Z_3 \times Z_3, U(1))$
\footnote{This is because $Z_3^n$ forms a vector space over the field $Z_3$.}.

\subsection{SPT ground state and a parent Hamiltonian}
Different symmetry-protected topological phases of $d$-dimensional spin systems are classified by the $(d+1)$-cocycles of their
symmetry group $G$.
Any nontrivial SPT state is obtained from the trivial product state by the action of a symmetric unitary operator
constructed from the local quantum gates (finite-depth quantum circuit). In the $d=2$ case, the unitary transformation
is diagonal in the Ising (Potts) basis and has the following form:
\begin{equation}
	U=\prod_{\Delta } \nu_3(0,n_1^\Delta,n_2^\Delta,n_3^\Delta)^{S(\Delta)}=\prod_{\Delta } \omega_3(n_1^\Delta,n_2^\Delta-n_1^\Delta,n_3^\Delta-n_2^\Delta)^{S(\Delta)},
\end{equation}
where $\nu_3=e^{i\varphi}$ is the three-cocycle in the invariant form containing one more argument \eqref{nu}, \eqref{dnu}, $n_1^\Delta, n_2^\Delta, n_3^\Delta$ are the vertices of the triangle $\Delta$ with the lower index indicating their color.
The first argument of all $\nu_3$ may be chosen as any other constant.
The $G$-invariance of $U$ follows then from the cocycle condition $\delta\nu_3=1$ as can be verified \cite{Yoshida-2017}.
Amazingly, for the $G=(\times Z_3)^3$ group and aforementioned symmetric cocycle \eqref{3-cocycle-1},
the above unitary map admits a simpler form via the $\omega$ description: $U=\prod_{\Delta=\langle i,j,k\rangle } \omega_3^{123}(n_i,n_j,n_k)^{S(\Delta)}$
as was shown for any cyclic group \cite{Wei-2018}.

For $G=Z_3$, we have found that the cohomology group consists of the cocycle\\ $\omega_3(n_1, n_2, n_3) = \epsilon^{n_1n_2n_3(n_1+n_2)}$, with its square and the trivial element. The unitary operator that gives the corresponding symmetry-protected state is then
\begin{equation}
	U=\prod_{\Delta}\omega_3 (n_1^\Delta, n_2^\Delta-n_1^\Delta, n_3^\Delta-n_2^\Delta)^{S(\Delta)} = \prod_{\Delta} \varepsilon^{S(\Delta) n_1n_2n_3(n_2-n_1)},
\end{equation}
where we cancel out terms of the exponent that don't contain all $n_i^{\Delta}, i=1,2,3$ with corresponding terms from the neighboring triangles. Starting with equivalent cocycles $\omega_3'=n_1n_2n_3(n_2+n_3)$ and $\omega_3'' = n_1n_2n_3(n_2-n_1-n_3)$, we will get the following unitary operators:
\begin{equation}
	\label{UU}
	U' = \prod_{\Delta} \varepsilon^{S(\Delta) n_1n_2n_3(n_3-n_2)}, \quad U''=\prod_{\Delta} \varepsilon^{S(\Delta) n_1n_2n_3(n_1-n_3)}
\end{equation}

As one can see, doing the cyclic permutation on the indices does not change the SPT phase. This should be expected as it corresponds to the rotational symmetry of the lattice around the center of one of the triangles. On the other hand, odd permutations change the orientation of the lattice and produce another nontrivial SPT phase that is complex conjugate to $\omega_3$.

\section{ $Z_3 \times Z_3 \times Z_3$ 
SPT paramagnet}

In this section, we generalize the construction of the $Z_2$ SPT version of the Ising paramagnet \cite{Levin-Gu} to the case of the $Z_3 \times Z_3 \times Z_3$ symmetry. We formulate models belonging to different SPT phases corresponding to basic elements of the cohomology group $H^3(Z_3\times Z_3 \times Z_3, U(1))=(\times Z_3)^7$. These bulk models differ by their ground state wave function, which, nevertheless, can be connected via $(\times Z_3)^3$ symmetric unitary transformation
\bea
\label{UT}
|0 \rangle_a= U_{ab}|0\rangle_b,\quad    a,b =1,2,\cdots, 7.
\ena
The unitary matrix, $U_{ab}$, however, can not be continuously transformed to identity via a line
of local unitary transformations  $U_{ab}(t)$,  such that $U_{ab}(1)=U_{ab}, \; U_{ab}(0)=1$ being   $(\times Z_3)^3$ symmetric all the way,
namely $ X^+_i U_{ab}(t)X^-_i=U_{ab}(t)$, see the definition \eqref{Ising-3}. In this section, following the approach of Ref.~\cite{Levin-Gu}, we
make a unitary transformation of spins by use of the cocycle $\omega^{123}_3$
described above \eqref{3-cocycle-1}. Further, we will consider lattice with boundary and find out edge state Hamiltonian, which should be gapless as an indicator of the nontrivial SPT phase. Other possible SPT states generated with the present approach and their boundary excitations will be analyzed in the subsequent publications.

Because the bulk and boundary parts of the paramagnetic Potts model (\ref{ham}) do not interact, their spectra are independent and can be calculated separately.
In the paramagnetic phase, the model has a ground state energy $E_{gr}=-2 \emph{Volume}(R)$ and a gap $\Delta_P=3$.
\def\MyColorList{{"white","gray","black"}}
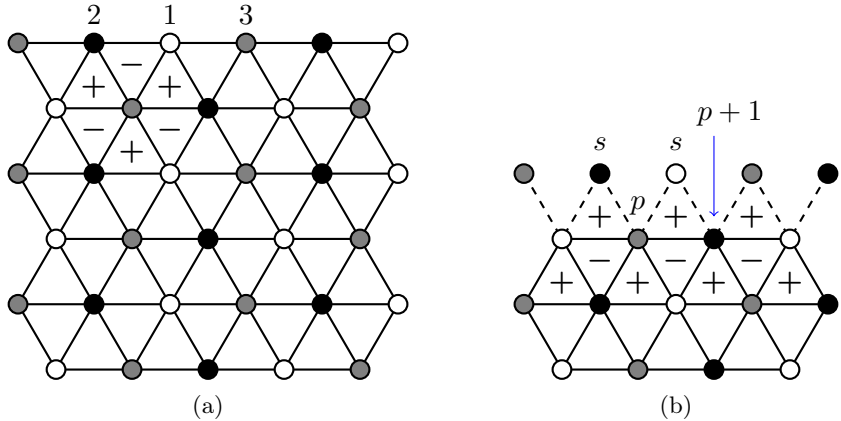
\begin{figure}[t!]
	\centering
	\subfloat[]{
		\begin{tikzpicture}
			\coordinate (p1) at ({\r+\dx},-\h);
			\coordinate (p2) at ({\r},0);
			\draw  node[above,yshift=3pt] at (p2) {$2$} node[above,yshift=3pt] at ($(p2)+(\r,0)$) {$1$}
			node[above,yshift=3pt] at ($(p2)+({2*\r},0)$) {$3$} ;
			\foreach \ang in {30,150,270}
			\node at ($(p1)+(\ang:2/3*\h)$) {$\bm{+}$};
			\foreach \ang in {90,210,330}
			\node at ($(p1)+(\ang:2/3*\h)$) {$\bm{-}$};
			\draw  (0,0)   --++  ({5*\dx},{-5*\h});
			\draw  (\r,0)   --++  ({5*\dx},{-5*\h});
			\draw  ({2*\r},0)   --++  ({5*\dx},{-5*\h});
			\draw  ({3*\r},0)   --++  ({4*\dx},{-4*\h});
			\draw  ({4*\r},0)   --++  ({2*\dx},{-2*\h});
			\draw  (0,{-2*\h})   --++  ({3*\dx},{-3*\h});
			\draw (0,{-4*\h})   --++  (\dx,-\h);
			\draw  ({3*\r},0)   --++  ({-5*\dx},{-5*\h});
			\draw  ({4*\r},0)   --++  ({-5*\dx},{-5*\h});
			\draw  ({5*\r},0)   --++  ({-5*\dx},{-5*\h});
			\draw  ({2*\r},0)   --++  ({-4*\dx},{-4*\h});
			\draw  ({\r},0)   --++  ({-2*\dx},{-2*\h});
			\draw  ({5*\r},{-2*\h})   --++  ({-3*\dx},{-3*\h});
			\draw  ({5*\r},{-4*\h})   --++  ({-\dx},{-\h});
			
			\draw  (0,0)   --++  (\r,0) ++  (\r,0)   --++  (\r,0) ++  (\r,0)   --++  (\r,0);
			\draw (\r,0)   --++  (\r,0)  ++ (\r,0)   --++  (\r,0);
			\draw  (\dx,-\h)   --++  (\r,0) ++  (\r,0)   --++  (\r,0);
			\draw ({\r+\dx},-\h)   --++  (\r,0)  ++ (\r,0)   --++  (\r,0);
			
			\draw (0,{-2*\h})   --++  (\r,0) ++  (\r,0)   --++  (\r,0) ++  (\r,0)   --++  (\r,0);
			\draw  (\r,{-2*\h})   --++  (\r,0)  ++ (\r,0)   --++  (\r,0);
			\draw (\dx,{-3*\h})   --++  (\r,0) ++  (\r,0)   --++  (\r,0);
			\draw  ({\r+\dx},{-3*\h})   --++  (\r,0)  ++ (\r,0)   --++  (\r,0);
			
			\draw  (0,{-4*\h})   --++  (\r,0) ++  (\r,0)   --++  (\r,0) ++  (\r,0)   --++  (\r,0);
			\draw (\r,{-4*\h})   --++  (\r,0)  ++ (\r,0)   --++  (\r,0);
			\draw  (\dx,{-5*\h})   --++  (\r,0) ++  (\r,0)   --++  (\r,0);
			\draw ({\r+\dx},{-5*\h})   --++  (\r,0)  ++ (\r,0)   --++  (\r,0);
			
			\foreach \ny in {0,...,2}
			\foreach \nx in {0,...,5} {
				\pgfmathparse{\MyColorList[mod(\nx+1,3)]}
				\edef\col{\pgfmathresult}
				\filldraw[fill=\col]({\nx*\r},{-\ny*2*\h})  \cir;
			}
			\foreach \ny in {0,...,2}
			\foreach \nx in {0,...,4}
			\pgfmathparse{\MyColorList[mod(\nx,3)]}
			\edef\col{\pgfmathresult}
			\filldraw[fill=\col] ({\nx*\r},{-\ny*2*\h}) ++ (\dx,-\h) \cir;
		\end{tikzpicture}
		\label{fig-3a}
	}
	\hspace{1cm}
	\subfloat[]{
		\begin{tikzpicture}
			\coordinate (p1) at  (\r,0);
			\coordinate (p2) at   ({2*\r},0);
			\coordinate (p3) at   ({\r+\dx},-\h);
			\coordinate (p4) at   ({2*\r+\dx},-\h);
			\coordinate (p5) at   ({2*\r+\dx},\dx);
			\draw  node [above,yshift=5pt] at (p1) {$s$}   node [above,yshift=5pt] at (p2) {$s$}
			node [above,yshift=5pt] at (p3) {$p$}     node (n5) [above,xshift=6pt] at (p5) {$p+1$};
			\draw[thin,blue, ->] (p5) -- ($(p4)+(0,\eps)$);
			\foreach \i in {1,2,3} {
				\node at ($(\i*\r,-2/3*\h)$) {$\bm{+}$};
				\node at ($(\i*\r,-4/3*\h)$) {$\bm{-}$};
			}
			\foreach \i in {0,1,2,3} {
				\node at ($(\i*\r+\dx,-5/3*\h)$) {$\bm{+}$};
			}
			\draw   (\dx,-\h)   --++  (\r,0)  ++ (\r,0)  --++ (\r,0);
			\draw  ({\dx+\r},-\h)   --++  (\r,0);
			\draw (0,{-2*\h})   --++  (\r,0)  ++ (\r,0)  --++ (\r,0) ;
			\draw (\r,{-2*\h})   --++  (\r,0)  ++ (\r,0)  --++ (\r,0) ;
			\draw (\dx,{-3*\h})   --++  (\r,0)  ++ (\r,0)  --++ (\r,0);
			\draw   ({\dx+\r},{-3*\h})   --++  (\r,0);
			\foreach \nx in {0,...,3}
			\draw[dashed] (\nx,0) -- ++ (-60:\r) -- ++ (60:\r);
			\foreach \nx in {0,...,3}{
				\draw ({\nx*\r+\dx},{-\h}) -- ++ (-60:\r);
				\draw ({\nx*\r},{-2*\h}) -- ++ (60:\r);
				\draw ({\nx*\r},{-2*\h}) -- ++ (-60:\r);
				\draw ({\nx*\r+\dx},{-3*\h}) -- ++ (60:\r);
			}
			\foreach \ny in {0,...,1}
			\foreach \nx in {0,...,4}
			\pgfmathparse{\MyColorList[mod(\nx+1,3)]}
			\edef\col{\pgfmathresult}
			\filldraw[fill=\col] ({\nx*\r},{-\ny*2*\h})  \cir;
			\foreach \ny in {0,...,1}
			\foreach \nx in {0,...,3}
			\pgfmathparse{\MyColorList[mod(\nx,3)]}
			\edef\col{\pgfmathresult}
			\filldraw[fill=\col] ({\nx*\r},{-\ny*2*\h}) ++ (\dx,-\h) \cir;
		\end{tikzpicture}
		\label{fig-3b}
	}
	\qquad
	\caption{\label{fig-3}
		{\bf a)}  Triangular lattice partitioned into three sublattices, each corresponding to a single $Z_3$ group in the threefold product.
		Related spins are shown, respectively, in white, gray, and black colors.
		The $\epsilon=\pm$ signs label the  "up" and "down"  triangles correspondingly.
		{\bf b)} The boundary of the SPT Potts model. The external (ghost) spins $\dots,s,s,\dots$ on the top are connected  with the boundary
		spins $\dots,p,p+1,\dots$ on the next line with the dashed bonds.
	}
\end{figure}
To this end, we apply a unitary transformation to the  bulk spins, $\bar{X}^{\pm}_{\bf r} = U X^{\pm}_{\bf r} U^{\dagger}$, related to
the nontrivial cohomology element $\omega_3^{123}$ with
the matrix
$U=\prod_{\langle p,q,r\rangle }\varepsilon^{ \epsilon  n^{(1)}_p n^{(2)}_q n^{(3)}_r}$.
Here the sites in $\langle p,q,r\rangle $ are vertexes of  lattice triangular faces, and
{the product is taken} over
all faces with staggered sign
$\epsilon = \pm 1$ as was shown in Fig.~\ref{fig-3a}.
This unitary transformation will have a larger $Z_3\times Z_3 \times Z_3$
symmetry, because one can check that $X^\pm_i U = U X^\pm_i$.
To see this symmetry
we use formula (\ref{QGalgebra}) for commuting a single  term of the  triangle
${\bf m}$ with the
unitary transformation of color 1,
namely
\bea
\label{sym-1}
\varepsilon^{\epsilon n^{(1)}_p n^{(2)}_q n^{(3)}_r}X_{{\bf m},1} =X_{{\bf m},1}\varepsilon^{\epsilon (n^{(1)}_p +1) n^{(2)}_q n^{(3)}_r},
\ena
where $p, q, r,$ are sites of the triangle ${\bf m}$. As we see, there are extra terms of the form $\varepsilon^{\epsilon  n^{(2)}_q n^{(3)}_r}$,  appearing after commutation.
But a similar term appears after commuting $X_1$ with
the neighbor triangle with the same $(q, r)$ link, which has an opposite
sign (see Fig.~\ref{fig-3a}). Therefore, both extra terms cancel each other after commutation
with $U$. This happens with any $X_i, \; i=1,2,3$, and thus the symmetry of this term is larger.

Since the unitary transformation
\bea
\label{U}
U=\prod_{\langle p,q,r\rangle }\varepsilon^{ \epsilon \; n^{(1)}_p n^{(2)}_q n^{(3)}_r}
\ena
is symmetric under Ising transformation $X^\pm_i$ with colors $i=1,2,3$ belonging to
$Z_3 \times Z_3 \times Z_3$, it creates a particular $Z_3$ SPT phase,
belonging to the family of the elements of the group $(\times Z_3)^7$.
The Hamiltonian of corresponding
phase in bulk reads
\bea
\label{H4}
\bar{H}_{bulk}=U^+ H_{bulk}U=- \sum_{({\bf m},i)\in R} (\bar{X}^+_{{\bf m},i}+\bar{X}^-_{{\bf m},i}),
\ena
where at each site $({\bf m},i)$ in bulk, we have
\bea
\label{barS}
\bar{X}^{\pm}_{{\bf m},i}=U X^{\pm}_{{\bf m},{i}}U^+= \sum_{\langle p,q\rangle ,i\neq j,k}{X}^{\pm}_{{\bf m},i} \varepsilon^{\mp \epsilon  n^{(j)}_p n^{(k)}_q },
\ena
and $\langle p,q\rangle $ are all neighboring sites on the hexagon containing site $({\bf m},i)$
in the middle, while $p$ and $q$ are $j$, $k$ color partners of the
$({\bf m},i)$.

Clearly, the algebra (\ref{QGalgebra}) is unaffected by the unitary transformation (\ref{U}). It remains as an algebra of symmetry of the Hamiltonian (\ref{H4}) of the new phase.
After unitary transformation, the Hamiltonian (\ref{H4}) in bulk has the same
spectrum as the original one and the system will remain intact as for the
paramagnet. Interestingly, the bulk Hamiltonian, $\bar{H}_{bulk}$, also has three-state permutation symmetry $S_3$. However,
the transformation matrix $U$ given by Eq.~(\ref{U}) does not preserve it.
One can check that under permutation transformations $S_{12}, S_{13}, S_{23}$ (\ref{S3})
the SPT phase is changing within the $Z_3$ subgroup of the group of cohomology,
while permutations $X^{\pm}$ leave the SPT phase intact.

\vspace{1cm}
\section{Symmetry protected edge states}

The situation is different on the boundary of the system. The transformation (\ref{U}) is
restricted to triangles inside the sample:
\bea
\label{U-1}
U_b=\prod_{\langle p,q,r\rangle \in R}\varepsilon^{ \epsilon \; n^{(1)}_p n^{(2)}_q n^{(3)}_r}.
\ena
This expression is incomplete to be symmetric since the boundary links missed the associated triangles outside of the sample $R$, see Fig.2b and Fig.3. Thus, $U_b$
can be expressed as a product of two parts $U_b = U_R U_{\partial R}$, with
\bea
\label{sim-violation}
U_R=\prod_{\langle p,q,r\rangle \in R-\partial R}\varepsilon^{ \epsilon \; n^{(1)}_p n^{(2)}_q n^{(3)}_r}, \quad &&
U_{\partial R}=\prod_{\langle q,r\rangle \in \partial R}\varepsilon^{ \epsilon \; n^{(1)}_p n^{(2)}_q n^{(3)}_r}.
\ena
Here $U_R$ contains all the triangular faces with links inside the bulk, $R$, while
$U_{\partial R}$ consists of the triangles which have at least one bond on the boundary
$\partial R$. It is clear, that $U_R$ is symmetric under Ising transformation, $X^+ U_R X^- = U_R$, while $U_{\partial R}$ is not:
\bea
\label{sim-violation-1}
X^+ U_{\partial R} X^- = U_{\partial R} \prod_{p\in \partial R}\varepsilon^{\epsilon_{p,p-1} \; n^{(i_{p-1})}_{p-1} n^{(i_{p})}_{p} + s_p  \; n^{(i_p)}_{p}} =  U_{\partial R} \;  \Omega.
\ena
Here we introduced a notation $i_p$ defining the sequence of color sites on the boundary. Symbol $\epsilon$ is used to represent
the sign of the triangle touching the boundary,  while $s_p$ is an index that depends on the number of triangles inside of $R$, associated with the point $p$, and also on their sign $\epsilon$. In Fig.3a and Fig.3b, we present examples of boundaries with a particular distribution of signs $\epsilon$ and corresponding $s_p$. One finds that when three triangles with staggered signs merge at the point $p$ inside of $R$ or one triangle is associated with $p$, then $s_p=\pm 1$. When two or four triangles are merging, then $s_p=0$. In the presented examples, one can see that the extra $\Omega$ term is equivalent to adding to $U_R$ a chain of external (outside of bulk $R$) triangles with fixed spin $-1$.
\begin{figure}[h]	
	\includegraphics[width=75mm]{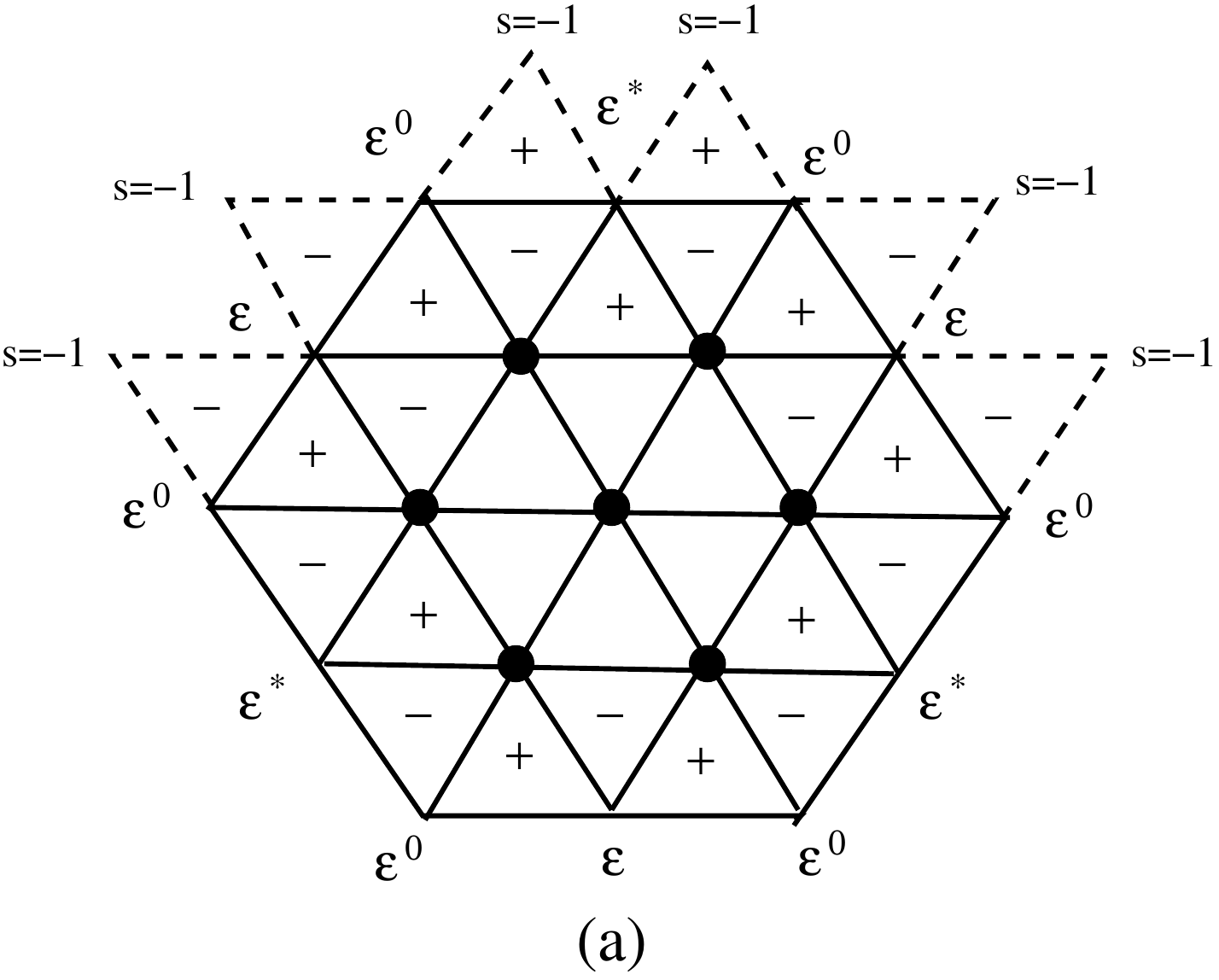} 	
	\label{fig-31a}
	\includegraphics[width=75mm]{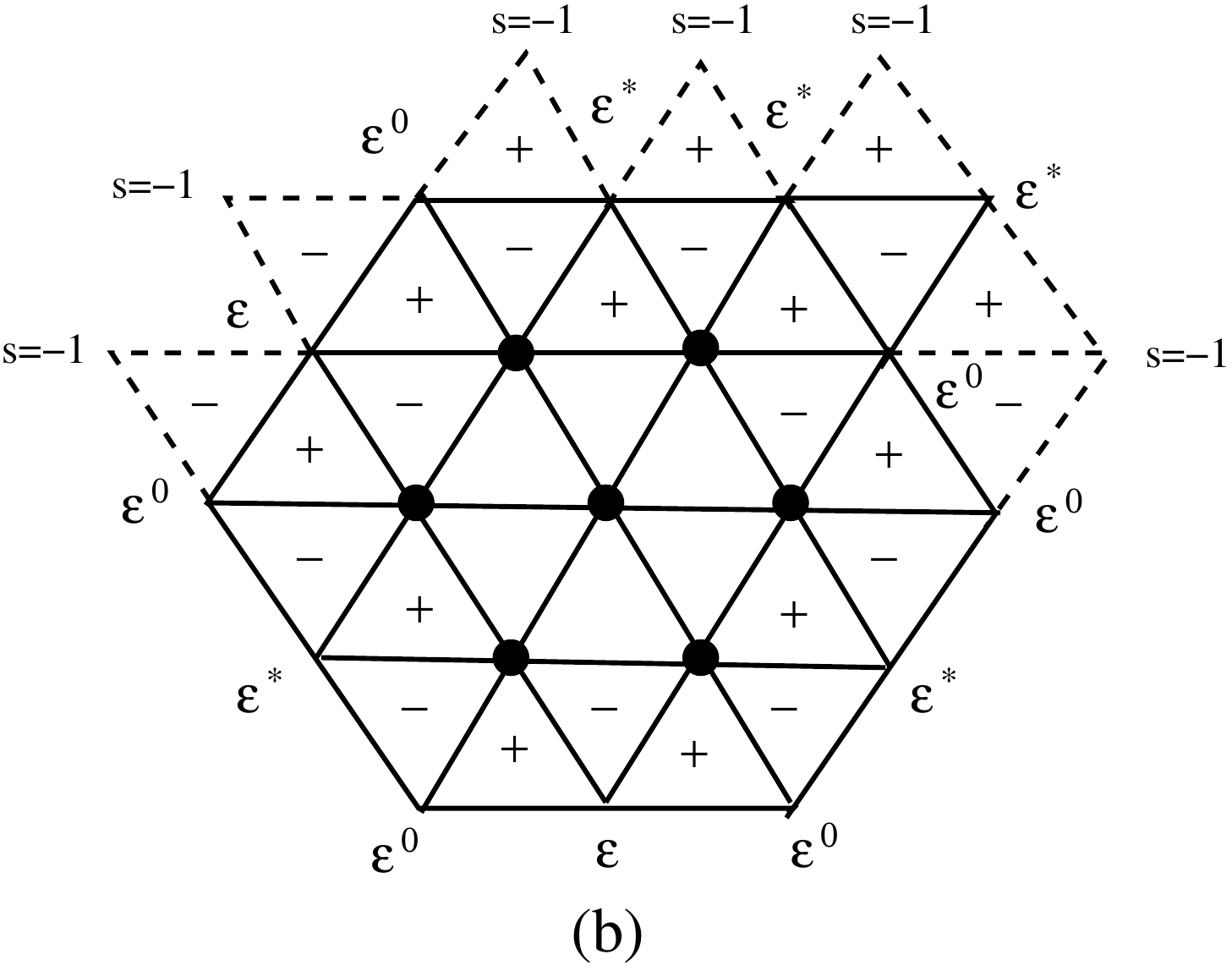}   	
	\label{fig-31b}
	\caption{Two configurations of boundaries. a) Length of the boundary is even.
		The product of $\varepsilon$-s on even and odd sites separately is 1.
		b) Length of the boundary is odd. The product of all $\varepsilon$-s on sites is 1. We present here an example of a boundary with a particular disposition of signs of triangles and phases $s_p$ corresponding to them. External fixed spins here are -1.  }
\end{figure}
Thereby, the unitary transformation (\ref{U-1}) acting on the boundary gives a non-symmetric result. Its application will create new boundary modes, which, being massless, will indicate the appearance of phase transition.

Since the map, $i_p$, of colors on boundary spins is identical (one-to-one), and higher symmetry is absent, we drop the color index from the spin variables from now on.
Application of $U_R$ on the bulk states gives modified $Z_3$ spins (\ref{H4}-\ref{barS}), while on a boundary, it yields
\bea
\label{boundary-X}
U_R^+ U_{\partial R}^+ X_p^{\pm} U_{\partial R}U_R= \bar{X}_p^{\pm}.
\ena
The unitary transformation $U_b$ of the original paramagnetic Potts model (\ref{ham})
gives a modified Hamiltonian
$\bar{H}=U_b^+ H U_b=\bar{H}_{bulk}+\bar{H}_{edge}$, where the bulk part is defined
by (\ref{H4}), and it is topologically nontrivial. The edge spins will give an edge term
\bea
\label{H4-edge}
\sum_{p\in \partial R}(\bar{X}^+_p + \bar{X}^{-}_p).
\ena

Let us now turn to analyze the symmetry property of the Hamiltonian $\bar H$. Its bulk part $\bar{H}_{bulk}=U_b^+ H_{bulk}U_b$ is symmetric  under $Z_3 \times Z_3 \times Z_3$ group. However, following (\ref{sim-violation-1}) and (\ref{boundary-X}), the boundary part gives
\bea
\label{sim-violation-2}
\sum_{p\in \partial R}(X^+ \bar{X}^+_p X^- &+& H.c.)=\sum_{p\in \partial R}\big[X^+ U_R^+ U_{\partial R}^+ X_p^{\pm} U_{\partial R}U_R X^- + H.c.\big]\nn\\
&=&\sum_{p\in \partial R} \big[\Omega^+ \bar{X}^+_p \Omega +H.c.\big],
\ena
with
\bea
\label{omega}
\Omega=\prod_{p\in \partial R}\varepsilon^{\epsilon_{p,p-1} \; n_{p-1} n_{p} + s_p  \; n_{p}}.
\ena
As indicated above, we have dropped color indexes.
Application of the Ising group elements $X^{\pm}$ for the second time yields a different boundary
term
\bea
\label{boundary-X2}
\sum_{p\in \partial R} \big[\Omega^{+2} \bar{X}^+_p \Omega^2 +H.c.\big]=\sum_{p\in \partial R} \big[\Omega^- \bar{X}^+_p \Omega^+ +H.c.\big]
\ena
The first term in the exponent of $\Omega$ can be interpreted as a "triangle" with external fixed
spin $s=-1$. See the example presented in Fig.3a and Fig.3b. In the case of $\Omega^2$, it can be interpreted as an external spin $s=1$, while for $\Omega^3=1$, as an external spin $s=0$.
Therefore, in order to apply a similar unitary transformation to the boundary spins $S_p$ (see Fig.~\ref{fig-3b}), we,
following the prescription outlined in Refs.~\cite{Wen-2012,Levin-Gu},
add lattice sites outside of the bulk (see dotted lines in  Figs.~\ref{fig-3b}, Fig.3a
and Fig3b) and fix external classical spins $s=1,0,-1$ on a line passing through those points.

After fixing the external boundary configuration of spins $\{s\}$, the application of extended unitary transformation $U_{p,s}$ from (\ref{U}) to the edge spins, $p$, produces
\bea
\label{external}
\bar{X}^{\pm}_{p,s,s_p}=\Omega^{+s} \bar{X}^\pm_p\Omega^s= \bar{X}^\pm_p\varepsilon^{\mp \epsilon_{p,p+1} s n_{p+1} \mp \epsilon_{p,p-1} s n_{p-1} \mp \epsilon_{p,p-1} s_p s},
\ena
in correspondence with (\ref{sim-violation-2}-\ref{boundary-X2}).
Then the expression (\ref{external}) will give
$\bar{X}^{\pm}_{p,0}=X^{\pm}_p$. Having now unitarily transformed spins on the boundary, we can define a new edge Hamiltonian with global $Z_3$ Ising symmetry. In the most general form, the symmetric Hamiltonian reads
\bea
\label{ham-2}
\bar{H}_{edge}=-\sum_{p\in \partial R; s=1,0,-1} \big(\bar{X}^+_{p,s,s_p}+\bar{X}^-_{p,s,s_p}\big),
\ena
where $s_p$ is the external spin at boundary point $p$ (see Fig.3). Each term in this sum is a unitary transformation of $H_{edge}$ from (\ref{ham}). However, their sum over $s_p=1,0,-1$, is not a unitary transformation of $H_{edge}$. Therefore the spectrum of edge Hamiltonian $\bar{H}_{edge}$ does not coincide with the spectrum of $H_{edge}$ and, if it is gapless, is $Z_3\times Z_3 \times Z_3$ analog of the boundary XX model of the Levin-Gu $Z_2$ Ising paramagnet model \cite{Levin-Gu}.
There are various options for choosing an external spin configuration $\{s\}$. In this paper, we will present two of them: (a) the case when all external spins are equal and take values $(1,1, \cdots 1), (0,0, \cdots 0)$ and $(-1,-1,\cdots,-1)$, and (b) external spins are staggered  and take value $\{s,s'\}=(1,-1,1,-1 \cdots), (0,0,\cdots), (-1,1,-1,1,\cdots)$. We will show that these two options are unitary equivalent.

\subsection{Edge Hamiltonian}
\label{sec:another}
We enumerate boundary spins  as $p =1 \cdots N $, where $N$ is the length of boundary $\partial R$. Below we drop the bar from the spins $\bar{X}_{p,s}$.  The non-zero spin-$s$ boundary spins can be obtained from the $s=0$ component upon application of $X^{\pm}_p$, also upon applying the unitary transformation following $\Omega_a$
\bea
\label{external-2}
\Omega_a=\prod_{p}X^{+}_p \cdot \Omega.
\ena
Here the presence of $\epsilon$ in the exponent indicates what kind of triangle is
attached to the boundary from outside. See Fig.~\ref{fig-3}.b, where $\epsilon = +1$ triangles are attached.  One can check that
\bea
\label{B1}
\Omega_a^{- 1}\bar{X}^{\pm}_{p,s} \Omega_a =\bar{X}^{\pm}_{p,s+1},
\ena
which, after taking $s=0$ and using $\bar{X}^{\pm}_{p,0}=X^{\pm}_p$, leads to
\bea
\label{a)}
\bar{X}^{\pm}_{p,1}&=&\Omega^{-1}_a X^{\pm}_{p}\Omega_a = \varepsilon^{s_p}\varepsilon^{\pm  n_{p-1}}X^{\pm}_p\varepsilon^{\pm  n_{p+1} }, \\
\bar{X}^{\pm}_{p,-1}&=&\Omega^{-2}_a X^{\pm}_{p}\Omega^2_a=\varepsilon^{-s_p}
\varepsilon^{\mp  n_{p-1}}X^{\pm}_p\varepsilon^{\mp  n_{p+1} }.\nn
\ena
Here $s_p$ are phase factors, shown in Fig.3.
We have made use of the $Z_3$ property of $\Omega$ operators, namely $\Omega_a^2=\Omega^{-1}_a$. From now on we will use notations $Z_p^\pm=\varepsilon^{\pm n_p}$, as well as consider
boundary triangles marked by a $+$ sign.

Using Eq.~(\ref{a)}) with $\epsilon =+1$, the edge Hamiltonian Eq.~(\ref{ham-2}) for the set of constant external spin configurations becomes
\begin{align}
	\label{ham-3}
	H^{(a)}_{edge}&=
	-\sum_p \big(X^+_p+ \Omega_a^{-1} X^+_p \Omega_a+ \Omega_a X^+_p \Omega_a^{-1}+H.c.\big)
	\\
	&=-\sum_p \Big(X^+_p+\varepsilon^{s_p} Z^+_{p-1}X^+_p Z^+_{p+1}+\varepsilon^{-s_p}Z^-_{p-1}X^+_p Z^-_{p+1} +H.c.\Big),
	\nn
\end{align}
One can find that phase factors $\varepsilon^{\pm s_p}$ can be accumulated into $Z^\pm_p$ as a trivial gauge field (in other words, these phase factors can be gauged out). The fact that the gauge field always is trivial follows from Fig.3. and Fig.4. The proof can be done in two stages.
First, let us consider any particular shape of the boundary. For example, as shown in Fig.3a, containing an even number of sites. We have straight lines where, according to
the disposition of the sign of triangles, we have $\varepsilon$ or $\varepsilon^+$.
One can see that each straight line of  $\varepsilon's$ has a parallel
line of $\varepsilon^+$ (see Fig.3).
As a gauge phase factor, we should accumulate phases into $Z$ components of spins at the even $Z_{2k}^\pm$  and odd  $Z_{2k+2}^\pm$ positions separately. We will have two independent circuits of phase factors, each between even-even or odd-odd
points respectively. Since we always have an equal amount of $\varepsilon's$
and $\varepsilon^+$, their products will be equal to one in each of the even and odd circuits, respectively. Hence gauge flux is zero, and phase factors can be
accumulated into the $Z^\pm_p$'s without affecting their $Z_3$ algebras.
If we have a boundary with an odd length, as in Fig.3b, then we have one circuit,
but again, the product of $\varepsilon's$ is one, and phase factors can be gauged
out.

The second stage is the observation that by adding to any boundary a triangle, as it was done
by the addition of a triangle on the boundary in Fig.3a and getting the boundary in Fig.3b, we are adding $\varepsilon^3=1$ or $\varepsilon^{+ 3}=1$, see Fig.4. Hence, modifying
the boundary by triangles, we always have zero flux gauge field, which
can be accumulated into the $Z_p$'s.

\begin{figure}[h]	\hspace{3 cm}
	\includegraphics[width=90mm]{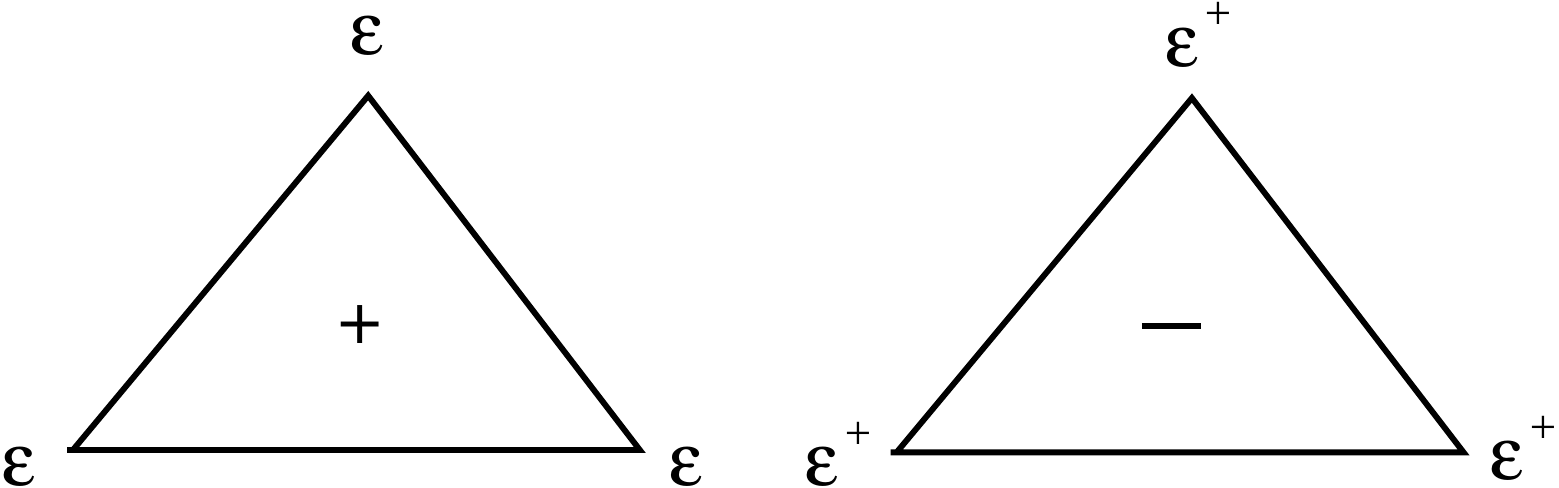} 	\label{fig-4}
	\caption{Disposition of $s_p$ phases from (\ref{sim-violation-1}) for triangles with + or - signs.}
\end{figure}

Therefore, from (\ref{ham-3}) we obtain following Hamiltonian
\begin{align}
	\label{ham-32}
	H^{(a)}_{edge}
	&=-\sum_p \Big(X^+_p+ Z^+_{p-1}X^+_p Z^+_{p+1}+Z^-_{p-1}X^+_p Z^-_{p+1} +H.c.\Big),
\end{align}

Besides translational invariance, the above Hamiltonian has global $S_3$ Ising symmetry. There are two integrals of motionAccording
\bea
\label{IM}
X=\prod_{i=1}^{N/4} X^+_{4i} X^+_{4 i+1} X^-_{4 i+2} X^-_{4 i+3},
\quad
P=\prod_{i=1}^{N} P^{13}_{i},
\ena
where $P^{13}$ is a permutation of spin $s=1$ and spin $s=-1$ components of $n$ in a basis where it is diagonal.
This operators are commuting with the edge Hamiltonian, $[H^{(a)}_{edge}, X]=0,\; [H^{(a)}_{edge}, P]=0$ and they are forming generators of $S_3$ (for details see the Appendix).
Moreover, the ground state wave function of the bulk model (\ref{ham}) in a basis of diagonal $n$ is $| V \rangle =(| 1\rangle +| 0\rangle +| -1\rangle)/\sqrt{3}$, which obviously invariant under permutations of $S_3$.

Importantly, the Hamiltonian $H^{(a)}_{edge}$ is self-dual under duality transformation
\bea
\label{dual}
X_p^{d\pm}=\Omega_a^{-1} X_p^\pm \Omega_a=Z_{p-1}^{\pm} X_p^\pm Z_{p+1}^{\pm},
\quad
Z_p^{d\pm}=Z_p^{\pm}
\ena
This transformation permutes three terms in the Hamiltonian (\ref{ham-3}), leaving its form invariant. Introducing three couplings in front of three terms in (\ref{ham-3}), we get the Hamiltonian
$$
H^{(a)}_{edge}(\lambda_1,\lambda_2,\lambda_3)=-\sum_p \big(\lambda_1 X^+_p+\lambda_2 Z^+_{p-1}X^+_p Z^+_{p+1}+\lambda_3 Z^-_{p-1}X^+_p Z^-_{p+1} +H.c.\big).
$$
It is straightforward to see that the transformation (\ref{dual}) permutes the position of three coupling constants. Also, taking into account the fact that the transformation (\ref{dual}) 
preserves the algebra (\ref{QGalgebra}), we arrive at the following relation on the spectrum of our Hamiltonian
\bea
\label{spectrum}
\lambda_1 E^{(a)}\Big(1,\frac{\lambda_2}{\lambda_1},\frac{\lambda_3}{\lambda_1}\Big)
=\lambda_2 E^{(a)}\Big(1, \frac{\lambda_3}{\lambda_2},\frac{\lambda_1}{\lambda_2}\Big)
=\lambda_3 E^{(a)}\Big(1, \frac{\lambda_1}{\lambda_3},\frac{\lambda_2}{\lambda_3}\Big)\nn
\ena
This property means that the transformation (\ref{dual}) is a duality transformation between different phases of the system. In contrast, the self-duality of the Hamiltonian at $\lambda_1=\lambda_2=\lambda_3$ indicates the presence of a quantum critical point. Moreover, the corresponding edge state is gapless, a hallmark of the SPT phase. According to the classification of SPT phases \cite{Wen-2012, Wen-2013}, SPT states of two-dimensional models are defined by the cohomology group $H^3(G, U(1))$, where G is the symmetry of the model. In our case, it is $Z_3$, and we get $H^3(Z_3, U(1))= Z_3$, indicating that our construction should support two different massless edge states that classify nontrivial SPT phases. The identity element of $Z_3$ gives trivial SPT
state, which does not have massless boundary states.

\subsection{Dual edge transformation with the $Z_3$ gauge field}
The concept of duality has proven to be a powerful tool for understanding the underlying structure of physical systems. The duality transformation involves interchanging the role of local degrees of freedom and interactions in a system. It has been used to explore various phenomena, from superconductivity to topological insulators. Here, we will focus on the 
 duality transformation for our edge Hamiltonian that can be constructed by involving the $Z_3$ gauge fields $\mu_{i,i+1}^{a}$. The logic of the construction is similar to the one presented in Ref.~\cite{Levin-Gu}
for the case of $Z_2$ symmetry.
The dual variables will correspond to the operators
$\check{X},\; \check{Z}$. Then
\bea
X_i= \check{Z}_{i-1}^{\dag} \check{Z}_{i}\mu_{i-1,i}^z, \qquad
Z_i=\mu_{i-1,i}^x.
\ena
Here $\mu_{i-1,i}^x\mu_{i-1,i}^z=\varepsilon \mu_{i-1,i}^z\mu_{i-1,i}^x$. The constraint of  gauge invariance
can be written as
\begin{equation}
	\mu_{i-1,i}^{x\dag}\mu_{i,i+1}^x \check{X}_i^{\dag}=1.
\end{equation}
Then
\begin{equation}
	\begin{aligned}
		Z_{i}^{\dag}Z_{i+1}&=\mu_{i-1,i}^{x\dag}\mu_{i,i+1}^x={\check{X}}_i,
		\\
		Z_{i-1}X_i Z_{i+1}
		&={\check{X}}_{i-1}^{\dag}\mu_{i-1,i}^x\check{Z}_{i-1}^{\dag} \check{Z}_{i}\mu_{i-1,i}^z\mu_{i-1,i}^x{\check{X}}_i=
		\check{Z}_{i-1}^{\dag} \check{Z}_{i}\mu_{i-1,i}^z\mu_{i-1,i}^{x\dag}{\check{X}}_{i-1}^{\dag}{\check{X}}_i,
		\\
		Z^{\dag}_{i-1} X_i Z^{\dag}_{i+1}&=
		{\check{X}}_{i-1}\mu_{i-1,i}^{x\dag}
		\check{Z}_{i-1}^{\dag}\check{Z}_{i}\mu_{i-1,i}^z\mu_{i-1,i}^{x\dag}{\check{X}}_i^{\dag}=\check{Z}_{i-1}^{\dag} \check{Z}_{i}\mu_{i-1,i}^z
		\mu_{i-1,i}^x{\check{X}}_{i-1}{\check{X}}_i^{\dag}.
	\end{aligned}
\end{equation}
The dual edge Hamiltonian in terms of new variables thus acquires the form:
\bea
{H}_{edge}^a=\sum_i \left( \check{Z}_{i-1}^{\dag} \mu_{i-1,i}^z\check{Z}_{i}+ \check{Z}_{i-1}\mu_{i-1,i}^{z\dag} \check{Z}_{i}^{\dag}\right)
\left(1+{\check{X}}_{i-1}\mu_{i-1,i}^{x}{\check{X}}^{\dag}_i+
{\check{X}}^{\dag}_{i-1}\mu_{i-1,i}^{x\dag}{\check{X}}_i\right).
\ena
This Hamiltonian represents a spin chain coupled to a $Z_3$ gauge field.

\subsection{Alternative description of the edge Hamiltonian}

The edge Hamiltonian $H^{(a)}_{edge}$, (\ref{ham-3}) can be modified to a simpler form. One can make a transformation $Z^{\pm}_{4 k+i}\rightarrow Z^{\mp}_{4 k+i},\; X^{\pm}_{4 k +i} \rightarrow X^{\mp}_{4 k +i},$ for $\; i=1,2 $, while keeping other spins intact. It is unitary
transformation with $S_{13}$ from (\ref{S3}) and preserves $Z_3$ algebra.
After this transformation, the Hamiltonian acquires the form
\begin{align}
	\label{ham-321}
	H^{(b)}_{edge}
	&=-\sum_p \Big(X^+_p+ Z^+_{p-1}X^+_p Z^-_{p+1}+Z^-_{p-1}X^+_p Z^+_{p+1} +H.c.\Big),
\end{align}
This edge model can be obtained from the staggered  external spin configuration, in which external spins take values $\{s,s'\}=(1,-1,1,-1 \cdots), (0,0,\cdots), (-1,1,-1,1,\cdots)$. Non-zero $s,s'$ components can be obtained from $s=0$ component $X^{\pm}_p$ by use of the following unitary transformation $\Omega_a$ applied to the boundary spins
\bea
\label{external-3}
\Omega_b=\prod_{p}X^{+}_p \cdot \prod_{p} \varepsilon^{ (-1)^{p} n_{p} n_{p+1}}.
\ena
Here we see that staggered external spins appear in the exponent of the $\Omega$. Due to the staggering of the external spins operator, $\Omega_b$ rotates neighboring spins in opposite directions
\bea
\label{B2}
\Omega_b^{-1}\bar{X}^{\pm}_{p,s,s'} \Omega_b =\bar{X}^{\pm}_{p,s-(-1)^p,s'-(-1)^p}.
\ena
Defining $\bar{X}^\pm_{p,0,0}=X^\pm_p$, the expression (\ref{B2}) gives
\bea
\label{b)}
\bar{X}_{p,1,-1}=\Omega^{-1}_b X_p^{\pm}\Omega_b.
\ena

As in the previous representation of the model, this transformation gives a unitary equivalent to bulk terms but not on the boundary. We take the sum of terms with different spin configurations for the edge Hamiltonian, and the sum is not equivalent to unitary transformation.
Therefore we obtained a new boundary term on top of the initial Hamiltonian, which reads
\bea
\label{ham-4}
H^{(b)}_{edge}&=&-
\sum_p \big(X^+_p+ \Omega_b^{-1} X^+_p \Omega_b+ \Omega_b X^+_p \Omega_b^{-1}+H.c.\big)
\\
&=&-\sum_p \Big(X^+_p+Z^+_{p-1}X^+_p Z^-_{p+1}+Z^-_{p-1}X^+_p Z^+_{p+1} +H.c\Big)\nn
\ena
This is another analog of the free fermion XX model edge term obtained by Levin and Gu for
$Z_2$ Ising case. This Hamiltonian is also translational invariant and is also self-dual under transformation
\bea
\label{dual-2}
X_p^{d\pm}=\Omega_b^{-1} X_p^\pm \Omega_b=Z_{p-1}^{\pm} X_p^\pm Z_{p+1}^{\mp},
\quad
Z_p^{d\pm}=Z_p^{\pm}.
\ena
Furthermore, it also fulfills the duality relation of the spectrum presented above for model (a).  Hence, this suggests that the present model also has massless boundary excitations determining the SPT phase. The detailed density-matrix renormalization group (DMRG) based simulation of the edge Hamiltonian is presented in the next section. We also numerically identify the universality class of the critical model.


\newcommand{\HTc}[1]{{\color[RGB]{176, 0, 239} #1}}

\section{Symmetries of the boundary model and its conformal properties}

\subsection{ 't Hooft anomaly}

As outlined in Sect.5, the boundary model with the Hamiltonian $H_{edge}^{(b)}$ in Eq.~(\ref{ham-4}) is invariant under the triality transformation Eq.~(\ref{b)}), which is  given by the  unitary
transformation  $\Omega_b$ defined in
Eq.~(\ref{external-3}). This suggests that the Hamiltonian at criticality is {\it{self-trial}}.

Now, we consider the semi-finite length realization of the operator $\Omega_b$ to detect the 't Hooft anomaly of the boundary model. Namely, let us consider 
\bea
\label{omega}
U_r(g)= X^{g +} \Lambda^g_r =\prod_{p=0}^r X_p^{g +} \cdot \prod_{p=0}^r \varepsilon^{(-1)^p g\;  n_p n_{p+1}}
\ena
where $g=0,1,2$ are the elements of the $Z_3$ group. These reduced operators, $U_r(g)$, no longer form a
$(Z_3)$ group, and instead are a projective representation of the symmetry group. The projective nature of this set of elements is reflected in the associativity condition
\begin{equation}
    U_r(g)(U_r(h)U_r(k))=\omega_3(g,h,k)(U_r(g)U_r(h)) U_r(k) \label{eq:projassoc},
\end{equation}
where $\omega_3(g,h,k)$ characterizes the nontrivial nature of the representation and
defines the 't Hooft anomaly of the $(Z_3)$ symmetry \cite{Nayak-2014, Levin-2021}. Projective factor $\omega_3(g,h,k)$ can be calculated directly and is equal $\omega_3(g,h,k)=\varepsilon^{-g h k}$, which is reflecting Eq.~(\ref{U}). 

It is important to emphasize that the product of terms in the expressions in Eqs.~(\ref{U}) and (\ref{omega}) can be split into even and odd parts, which are independent symmetry operators. Each of those acts on half of the degrees of freedom of the chain and expresses the presence of the chiral structure in the model.

\subsection{Winding symmetry and laterality}
  Besides triality the boundary model has  two additional symmetries that are interrelated for this particular edge Hamiltonian.  First one is connected with the topological winding number operator ${\cal W}$, defined for the $Z_3$ symmetric chain as


\begin{equation}
	\label{winding}
	{\cal W} = \frac{1}{3} \sum_{p} w_{p, p+1} , \hspace{0.5cm}
	{\text{where}}\hspace{0.5cm} w_{p, p+1} = \frac{i}{\sqrt{3}} (Z_{p} Z_{p+1}^\dagger - Z_{p}^\dagger Z_{p+1}).
\end{equation}
The commutativity $[H_{edge}^{(b)}, {\cal W}] = 0$ is straightforward to check, implying that the eigenstates of the Hamiltonian have a characteristic winding number.
\begin{figure}[h]	
	\centering
	\includegraphics[width=110mm]{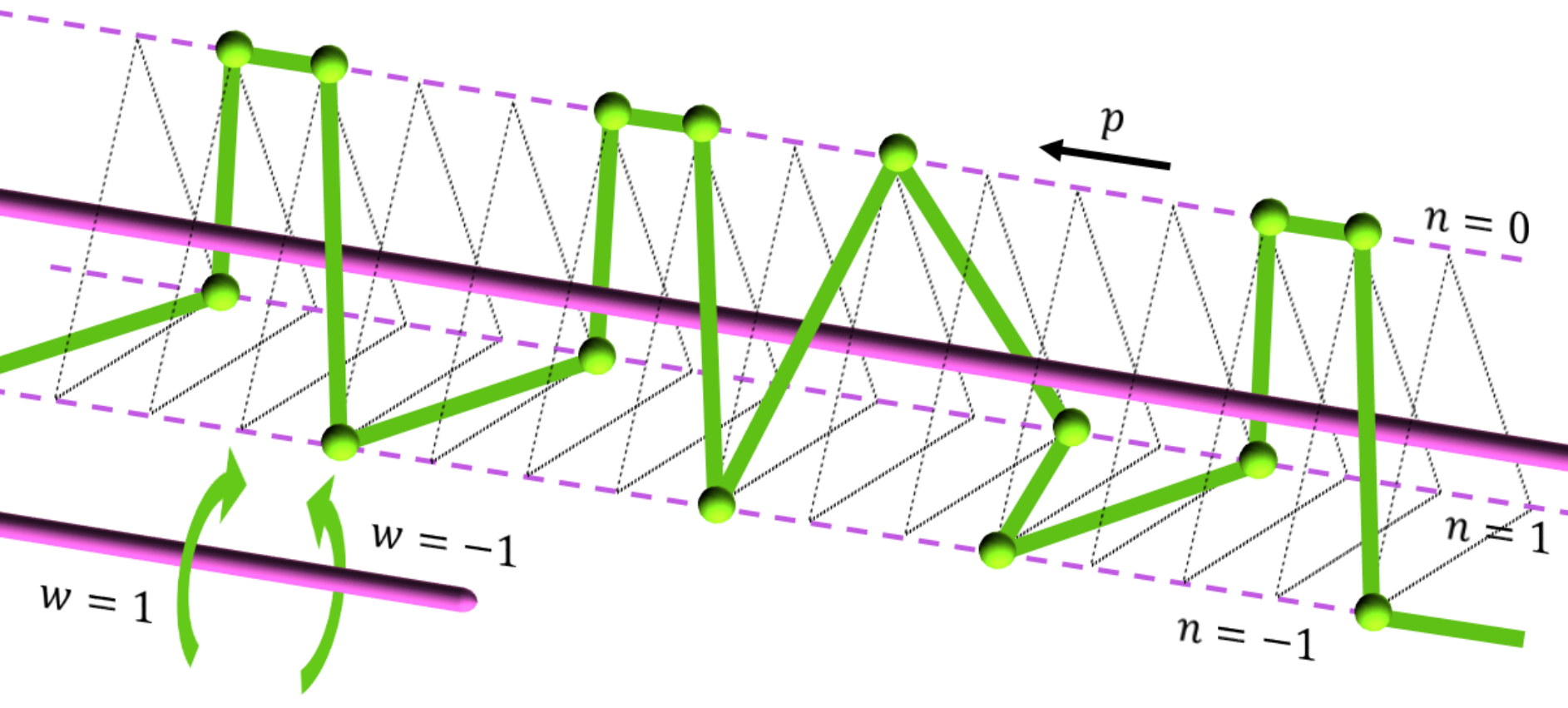}
	\caption{ Graphic representation of the concept of winding number
		in space $Z_3 \times \partial R$.
		The central line (thick purple) represents the edge,
		the dashed purple lines are the possible states $n=\{-1,0, 1\}$.
		Green balls indicate the states of each node.
		The winding number ${\cal W}$ is the number of full turns
		of the green polyline connecting neighboring node states
		around the central line.}
	\label{fig-wnum}
\end{figure}

Now let us gain an intuition on the winding symmetry and what ${\cal W}$ measures. For this purpose, we switch to the basis of eigenstates of $n_p=diag (-1,0,1)_p$ operators corresponding to the cite $p$. Then the eigenvalue of the operator
$w_{p, p+1}$ will be equal to $0$, for $n_{p + 1} = n_p$, while the eigenvalue 
$w_{p, p+1} = \pm 1$ for $n_{p + 1} \equiv n_p \pm 1 \mod 3$ respectively.
In other words, $w_{p, p+1}$ counts every $2i\pi/3$ phase 
of the value of the $Z_p$ in the complex plane
when moving from point $p$ to point $p+1$ of the 1D lattice (see Fig.\ref{fig-wnum} for the real space visualization of the winding number).
As the phase of $Z_p$ eventually returns to its starting value
after completing the full circle, eigenvalues of ${\cal W}$ can be understood as integers showing the number of full turns of $Z_p$ in the complex plane
during the traversal along the edge. This is the total winding number characteristic of a state under consideration.
Obviously ${\cal W}\in [-N/3, N/3]$.
This parameter itself has topological origins
if measured over space $of Z_3 \times \partial R$.

Another way to understand the commutativity $[H_{edge}^{(b)}, {\cal W}] = 0$ 
is to directly observe the action of the Hamiltonian
on eigenstates of $ n_p$, namely to states $\inb|{n_p}>$.
One may notice, that the action of $p$ component of $H_{edge}^{(b)}$
produces a non-zero result if and only if $n_{p-1} = n_{p+1}$.
Following this observation, one may rewrite the  Hamiltonian as
\begin{equation}
\label{subst}
	H_{edge}^{(b)} = -3 \sum_p (X_p + X_p^\dagger) \delta_{n_{p+1}, n_{p-1}}
\end{equation}
where $\delta$ is the Kronecker delta symbol.
The action of a component $p$ of the Hamiltonian substitutes
$\inb|{n, n_p, n}> \rightarrow -3 (\inb|{n, n_p - 1, n}> + \inb|{n, n_p + 1, n}>)$,
where $n_p\pm 1$ is assumed to be counted within $\mod 3$. From this representation of the Hamiltonian, it is clear that the eigenstates of it are also eigenstates of the winding number operator, ${\cal W}$, and thus characterized by the corresponding quantum number.
Throughout the rest of this paper, we refer to this symmetry as ${\cal W}$-symmetry.

We define the second symmetry, which is also written via local winding operators:
\begin{equation}
\label{later}
	{\cal L} = \sum_p (-1)^p w_{p, p+1}^2,
\end{equation}
which we call laterality and for the associated symmetry we will use a shortened notation ${\cal L}$-symmetry.
The corresponding eigenvalues also take integer values in the $ [-N/2, N/2]$ range.
Indeed, the substitution outlined below Eq.~(\ref{subst}) cannot possibly change ${\cal L}$,
as $w_{p-1, p}^2 = w_{p, p+1}^2$ both before and after it. This 
implies the commutativity $[H_{edge}^{(b)},{\cal L}] = 0$ for the Hamiltonian given in the form of Eq.~(\ref{subst}). The commutativity with the edge Hamiltonian in its original representation can also be checked directly.

	The winding number operator \eqref{winding} is commuting with translation operator $T$ of the chain while changing the parity under lattice reflection $P$: $[{\cal W}, T]=\{{\cal W}, P\}=0$. So, the eigenfunctions of the edge Hamiltonian \eqref{ham-32} always come in pairs with opposite winding quantum numbers $\pm w$ provided that $w\ne 0$. Each state of the pair with finite $w$ is mapped to the other state with $-w$ under the action of the $P$. Therefore, the nondegeneracy of the eigenstate implies trivial winding, $w= 0$.
 A similar logic applies to ${\cal L}$, but in this case, parity transformation does not change it,
and translation changes the sign: $[{\cal L}, P]=\{{\cal L}, T\}=0$.
The same implications for main state non-degeneracy can be made here,
which results in trivial laterality $l = 0$ for the main state.
	Our exact diagonalization studies of the edge Hamiltonian show that the ground state of even-site chains is nondegenerate. Hence, it must have the zero winding quantum number. For the odd $N$, the exact diagonalization of small chains in the sectors with specific momentum quantum numbers reveals a threefold degenerate lowest state with $w=0$ followed by the twofold degenerate excitation with opposite quantum numbers $w=\pm1$.

According to N\"others theorem any conserved quantity $Q = \sum_{q} q_p$ on 1D boundary creates current. If 
\bea
\label{KM-1}
i [H, q_p]=\nabla_p m_p, 
\ena
where $\nabla_p$ stands for a discrete derivative by index $p$,
then U(1) current $J^\mu = \sum_p j^\mu_p$, $\mu=0,1,$ with  $ j^0_p= q_p, j^1_p=m_p$ emerges, which fulfills conservation law $ \partial_\mu J^\mu =0 $,  with $\partial_0 = \partial_t$ and $\partial_1 = -\nabla_p$.
Our aim here is to make the relation (\ref{KM-1}) explicit for our ${\cal W}, {\cal L}$ symmetries with
$q^{\cal W}_p=\frac{1}{3}(w_{p, p+1}+w_{p-1, p}) $ and $q^{\cal L}_p=(-1)^{p+1} w^2_{p, p+1}+ (-1)^p w^2_{p-1, p}$ and find the corresponding $m^{\cal W}_p$ and $m^{\cal L}_p$.
The abovementioned commutators can be calculated straightforwardly and lead to
\bea
\label{Hw}
 &&i[H, q^{\cal W}_p] = \frac{i}{3}[ H, w_{p,p+1}+w_{p-1,p} ] = m^{\cal W}_{p+1}-m^{\cal W}_{p-1},\\
 \label{Hw1}
 &&i[H, q^{\cal L}_p] =  i[ H, (-1)^{p+1} w^2_{p,p+1}+(-1)^p w^2_{p-1,p}] = m^{\cal L}_{p+1}-m^{\cal L}_{p-1},
\end{eqnarray}
where
\bea
\label{Hm}
    m_p^{\cal W} &= i \delta_{n_{p-1}-n_{p+1}} X_p (3 \delta_{n_{p-1}-n_{p}-1}-1) +\textbf{h.c.}\\
    \label{Hm1}
    m_p^{\cal L} &= i \delta_{n_{p-1}-n_{p+1}} (-1)^p 
        \inb[{3\delta_{n_{p-1}-n_{p}}, X_p}] +\textbf{h.c.}\nn\\
        &=i (-1)^p \{w_{p-1,p}, m_p^{\cal W}\} +\textbf{h.c.}
\ena
In the process we have used identities $w_{p-1,p} = \delta_{n_{p-1} - n_p - 1} - \delta_{n_{p-1} - n_p + 1}$ and $w_{p-1,p}^2 = 1 - \delta_{n_{p-1} - n_p}$. Due to the presence of
$\delta_{n_{p-1}-n_{p+1}}$ terms in the expression of $m^{\cal W/L}$ we have substituted $n_p \rightarrow n_{p-2}$ in some terms of each commutator and derive the expressions (\ref{Hm}) for each symmetry.  Eq,~\ref{Hm1} indicates that ${\cal L}$ current depends on ${\cal W}$ current, and  we have {\em one single} independent $U(1)$ symmetry.
It is also straightforward to check, that $M^{\cal W} = \sum_p m_p^{\cal W}$ and
$M^{\cal L} = \sum_p m_p^{\cal L}$ obey commutativity relations
$[M^{\cal W}, {\cal W}] = [M^{\cal L},{\cal L}] = 0$.

\subsubsection{Associated $U(1)$ symmetry and the current algebra}

Now the question arises whether this symmetry creates $U(1)$ Kac-Moody  algebra or not \cite{Wang-Zou-2022} namely, whether there are holomorphic commutative chiral currents $j^{\pm}_{p} = (q_p \pm m_p)$ in the thermodynamic limit or not.
In case there are, they form a CFT with $U(1)$ Kac-Moody algebra.  Hence, to form a Kac-Moody algebra, currents must first fulfill the following
commutation relations in thermodynamic limit ($N \rightarrow \infty$ ) for low energy excitations:
\bea
\label{KM-2}
[j^{{\cal W/L},-}_p,j^{{\cal W/L},+}_{p'}]=0, \;\;\quad [j^{{\cal L},\pm}_p,j^{{\cal W},\pm}_{p'}]=0.
\ena
We show numerically that the charges of these currents commute, while the mentioned commutators decrease as $N^{-1.6}$ and $N^{-1.75}$ respectively for low-energy states. This means that we 
have commuting $U(1)$ left and right currents in the thermodynamic limit and low energy states.

 Secondly, for the currents to be holomorphic it is necessary to have
\bea
\label{KM-3}
\partial_{-}{j^+ }=\partial_+{j^- }=0,
\ena
with $\partial_\pm = \partial_t \pm \nabla_p$.
Notice that $\partial_{-}{j^+ }+\partial_+{j^- }=2\partial_\mu j^\mu = 0$.
The remaining condition $\partial_{-}{j^+ }-\partial_+{j^- } = 0$
on a lattice chain becomes
\begin{equation}
\label{KM-41}
i[H, m^{\cal W/L}_p]-(q^{\cal W/L}_{p+1}-q^{\cal W/L}_{p-1})=0.
\end{equation}

\begin{figure}[t]	
	\centering
	\includegraphics[width=110mm]{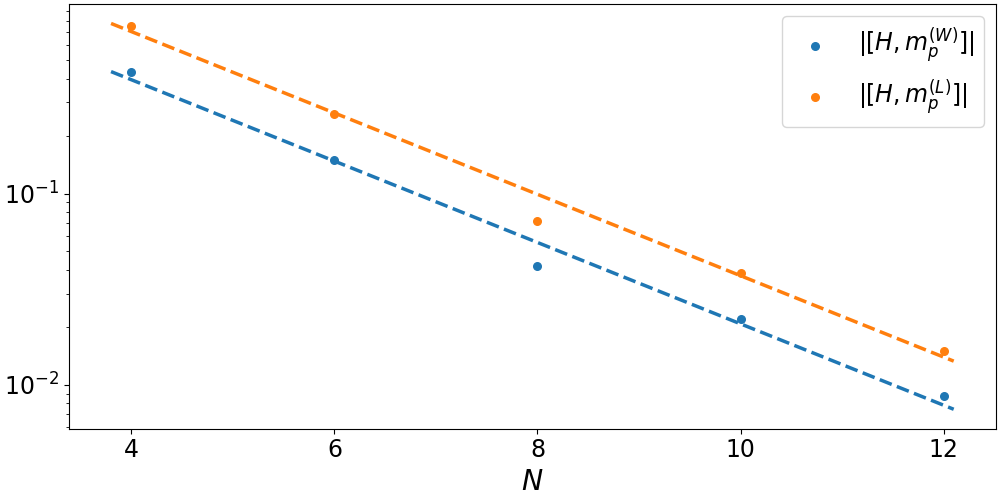}
	\caption{The behavior of the commutator term in the holomorphic factorization condition
 (\ref{KM-41}): exponential decrease upon increasing $N$. The vertical axis is in logarithmic scale.}
	\label{fig-7}
\end{figure}

Numerical calculations made on the basis of the exact diagonalization technique for chain sizes $N=4,6,8,10,12$ presented in Fig.\ref{fig-7} show, that  the commutator term
decreases with $N$ as $e^{-N/2}$ and goes to zero in the thermodynamic limit while
the second term in the  left side of the equation (\ref{KM-41}) is equal to zero.
 Therefore the condition (\ref{KM-41}) is fulfilled and we have holomorphic factorization
 in thermodynamic limit. Consequently, this generates Kac-Moody algebra for currents.

\begin{figure}[t]	
	\centering
	\includegraphics[width=110mm]{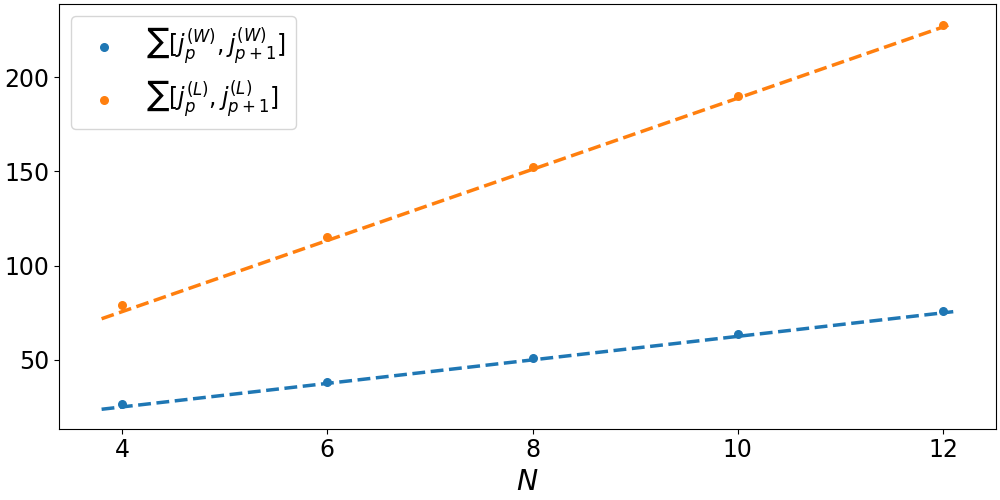}
	\caption{The behavior of $\sum_p [j^{\cal W/L,+}_p,j^{\cal W/L,+}_{p+1}]$ versus N. 
 Coefficients of linear functions are $\pi k_{\cal W} \simeq 2\pi$, $\pi k_{\cal L} \simeq 6 \pi$.
 Calculations have been made via the exact diagonalization technique.}
\label{fig-8}
\end{figure}

 Now, in order to find out whether this $U(1)$ symmetry is anomalous or not, we should calculate the commutators of currents $[j^{\cal W/L,\pm}_n, j^{\cal W/L,\pm}_{-n}]$
in the momentum space:
$j^{\cal W/L,\pm}_n = \frac{1}{2\pi}\sum_{p=1,N} e^{\frac{\pi i n p}{N}} j^{\cal W/L,\pm}_p$.
As $[j^{\cal W/L,\pm}_p,j^{\cal W/L,\pm}_{p'}] = 0$. For non-adjacent points, the relation transforms to
\bea
\label{KM-310}
[j^{\cal W/L,+}_n, j^{\cal W/L,+}_{-n}]= \frac{2i}{4 \pi^2} \sin\inb[{\frac{2 \pi n}{N}}] \sum_p 
 [j^{\cal W/L,+}_p,j^{\cal W/L,+}_{p+1}]+\frac{1}{4\pi^2}\sum_p [j^{\cal W/L,+}_p,j^{\cal W/L,+}_{p}],\nonumber\\
\ena
which is expected to be equal $ n k_{\cal W/L}$, with $k_{\cal W/L}$ being the level of Kac-Moody algebra. Our numerical calculations show that the second term in Eq.~(\ref{KM-310}) is zero, while the first therm $\sum_p[j^{\cal W/L,+}_p,j^{\cal W/L,+}_{p+1}]\simeq N \pi k_{\cal W/L}$ 
with $k_{\cal W} \simeq 2$, $k_{\cal L} \simeq 6$. The simulation results are presented in Fig.\ref{fig-8}.
This result suggests that 
\bea
\label{KM-4}
[j^{\cal W/L,+}_n, j^{\cal W/L,+}_{-n}] = n\; k_{\cal W/L}
\ena
in the thermodynamic limit, indicating that $U(1)$ symmetries  have  are anomalous. 
Numerical calculations presented in Fig.\ref{fig-8} were also made using the exact diagonalization technique for chain sizes $N=4,6,8,10,12$. 

It is necessary to emphasize that because the abelian  $U(1)$ current is defined up to a numerical coefficient, the level of the anomaly, $k_{\cal W/L}$, depends on current normalization.

\subsection{Gapless excitations and their conformal dimension}

\begin{figure}[t]	
	\centering
	\includegraphics[width=95mm]{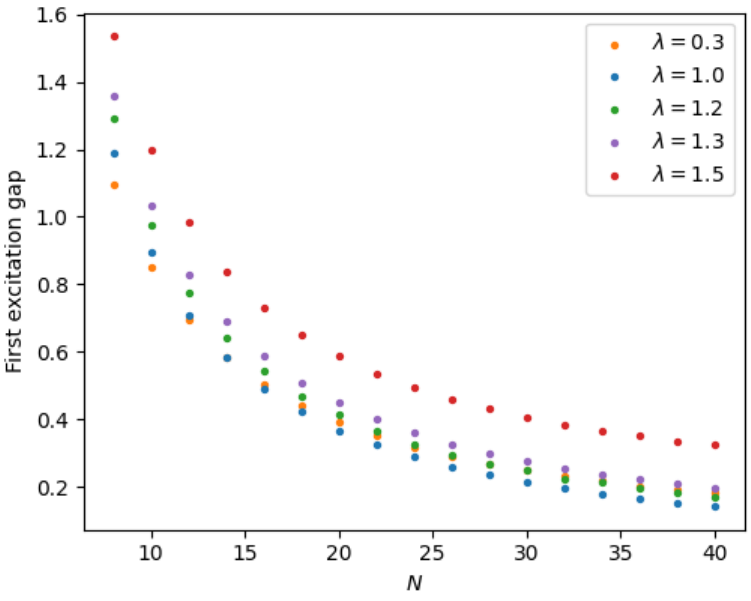}
	\caption{\label{fig-2} The density-matrix renormalization group (DMRG) based simulation of the excitation energy gap of the edge modes plotted versus the length of the boundary, $N$. The blue points correspond to the self-dual case with $\lambda=1$. The red points correspond to $\lambda=1.5$, green points to $\lambda=1.2$, purple points to $\lambda=1.3$, and orange points to $\lambda=0.3$. The extrapolation of the result suggests that the model with $\lambda=1$ in the $N\rightarrow \infty$ limit is critical. }
\end{figure}
\begin{figure}[t]	
	\centering
	\includegraphics[width=95mm]{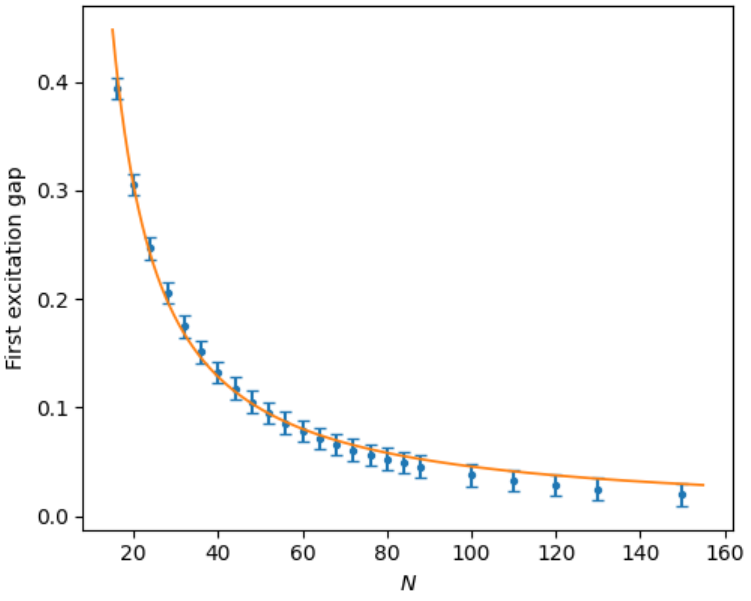}
	\caption{\label{fig-gap-5}The density-matrix renormalization group (DMRG) based simulation of the excitation energy gap of the edge modes at $\lambda=1$ is plotted versus the boundary length up to $N=150$. Error bars demonstrate the achieved precision of $10^{-2}$.}
\end{figure}

In order to see if the derived boundary Hamiltonian supports gapless excitations and conformal symmetry, we will start with the analysis of the spectrum of the low-energy excitations. The result of our DMRG simulations of the edge Hamiltonian presents strong evidence that this edge theory is indeed
gapless. The DMRG study of closed, periodic chains corresponding to our edge model required much more computational resources than open chains. This is because the time and capacity for calculating a given length and achieving the desirable precision on closed chains are much larger than on open chains. Therefore, we calculated open chains to get high-precision results using our present computing infrastructure.  
Fig.~\ref{fig-2} shows the excitation gap, namely the difference between the energy of the first excited state and the ground state plotted
versus the chain size $N$ for various values of $\lambda$.
Our simulation shows that the finite-size gap steadily decreases upon increasing $N$. 
 In particular, Fig.\ref{fig-2} shows that the lowest position among various curves has the curve with $\lambda=\frac{\lambda_2}{\lambda_1} =\frac{\lambda_3}{\lambda_1}=1$, which suggests that upon {\em departing} from the critical self-dual point, we are getting a finite excitation gap and thus entering into a massive phase.
The appropriate fit for the lowest curve simulated at the critical point with $\lambda=1$ presented in Fig.\ref{fig-gap-5} with a precision $10^{-2}$ yields
\bea
\label{numerics}
\Delta_N=  \frac{2 \pi\; 0.66625}{N} + {\cal O}(10^{-2}).
\ena 
	Here, according to Refs.~\cite{Cardy-1984,Cardy-1986}, the number $x_N={0.66625} \simeq \frac{2}{3}$ is the conformal dimension
	of the scaling operator concerned with the correlation length. In the next three subsections, we will present a detailed discussion of the corresponding primary fields with such dimensions in the conformal field theory that corresponds to the thermodynamic limit of our edge Hamiltonian. 

\subsection{Entanglement entropy}

Entanglement entropy (the Von Neumann entropy of the reduced density matrix for the subsystem), which measures the degree of entanglement between subsystems of a larger quantum system, has become an increasingly important area of research in recent years\cite{Huerta1, Huerta2, Huerta3}. In particular, the study of entanglement entropy in one-dimensional quantum chains has received significant attention for its relevance to condensed matter physics and potential applications in quantum information theory.
The entanglement entropy of a 1D open critical chain 
described by a 2D conformal field theory with the central charge $c$, can be found analytically\cite{Calabrese-2009}  and is given by 
$S_N(l)= a + \frac{c}{6}\log\Big(\frac{N}{\pi} \sin\Big[\frac{\pi l}{N}\Big]\Big)$,
where $a$ is a non-universal constant, $l$ is the subsystem length, and $N$ is the system size. For a closed chain, the coefficient in front of the $\log$ is twice larger (instead of $c/6$, it equals $c/3$). This formula describes the finite-size behavior of the entanglement entropy of the spin chain model at criticality.
It also gives a possibility to numerically evaluate the central charge from the finite-size scaling properties of a model. 

\begin{figure}[t]	
	\centering
	\includegraphics[width=85mm]{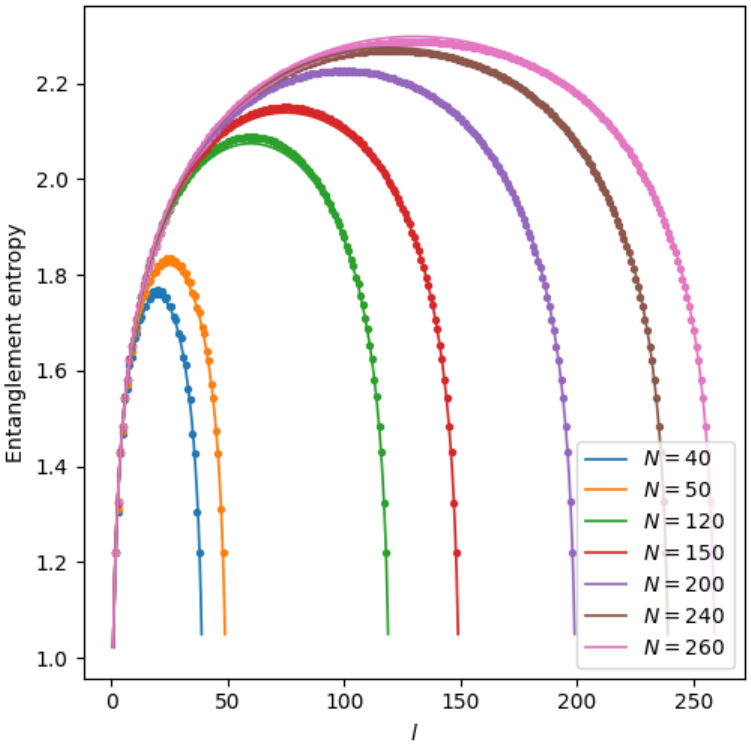}
	\caption{\label{fig-entanglement} Finite size entanglement entropy for
	open chain sizes $ N = 40, 50, 120, 150, 200, 240, 260$. The dots present numerical DMRG calculations for open chains using the ITensor package, while curves give the value of entanglement in conformal field theory with central charge $c=1.73$ and constant $a=1.021175$, upon fitting with the analytical expression for $S_N(l)$. }
\end{figure}

\begin{figure}[h]
	\centering
	\includegraphics[width=85mm]{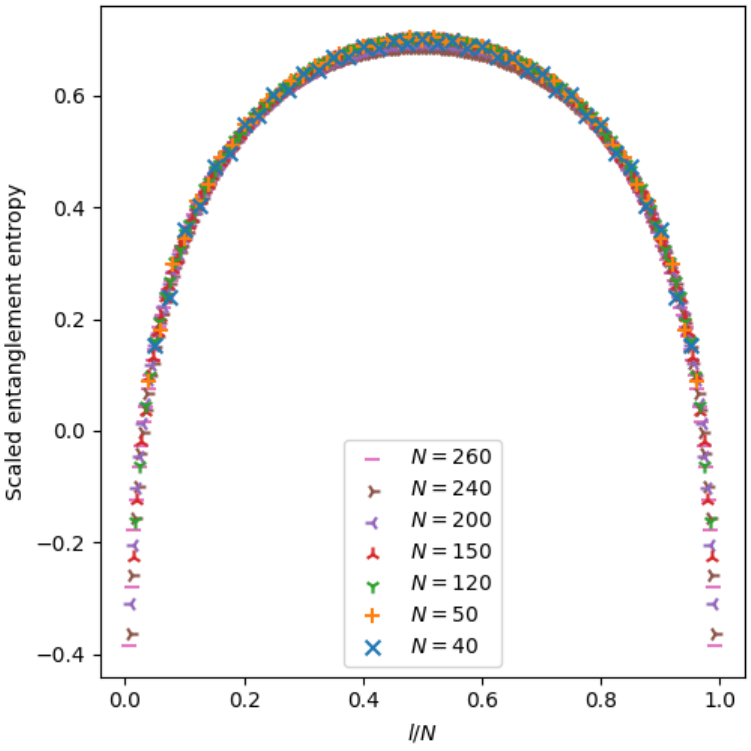}
	\caption{\label{fig-scaling} The scaled entanglement entropy of open chains
		$S_N-\frac{c}{6}\log(N)$ for sizes $ N = 40, 50, 120, 150, 200, 240, 260$  as a function of $\frac{l}{N}$.  }
\end{figure}

Here we will follow the aforementioned strategy to estimate the central charge of massless edge excitations. To this end, we have analyzed the entanglement entropy of the ground state using the iTensor package. Our numerical calculations are presented in Fig.\ref{fig-entanglement} and Fig.\ref{fig-scaling} for the chain lengths $N= 40, 50, 120, 150, 200, 240, 260$.  
  Fig.~\ref{fig-entanglement} shows numerical values for the entanglement entropy at various system sizes fitted by the analytic formula for $S_N(l)$. The precision fit yields a numerical estimate for the central charge $c=1.73$, while the non-universal constant is $a=1.02175$.   Similarly, Fig.~\ref{fig-scaling}  demonstrates the collection of all numerical points lying  on a single universal scaling function of a variable $l/N$ with the above values of $c$ and $a$: 
\bea
\label{scaled}
S_{scaled}(x)=S_N(l)-\frac{c}{6} \log[N]= a+ \frac{c}{6}\log\Big(\frac{1}{\pi} \sin[\pi x]\Big), \;\;  x=\frac{l}{N}, \;\; l= 2,\cdots N-1.
\ena
The iterations in our DMRG studies were made with precision $10^{-5}$, and our estimated error is $O(10)^{-2}$ for the excitation gap and $O(10)^{-3}$ for the central charge. This precision was set by our present computing infrastructure, but it can be verified and improved upon in subsequent studies. 

\section{Discussion and perspectives of the low energy theory of the edge Hamiltonian of the  $Z_3\times Z_3\times Z_3$ SPT paramagnet}

\subsection{Na\"ive search for matching conformal properties}

Having an estimate for the central charge, we now address the question: what conformal field theory will determine
the universal properties of the edge excitations of our $(\times Z_3)^3$ SPT paramagnet? 
 Since the cohomology group is $(\times Z_3)^7$, in principle, different phases associated with each non-identity element of this group may have different edge states described by different critical lattice Hamiltonians. We concentrate here on the edge of the phase (\ref{U-1}) defined by the $Z_3$ symmetry. 
Our numerical simulations show that the central charge $c\simeq 1.73$ and the conformal dimension of the lightest primary field is $\Delta=2/3$. Our main goal here is to understand what kind of CFT with appropriate t'Hooft anomaly \cite{t'Hooft} determines the low-energy limit of the edge model. The uncovered (while hidden from outset) $U(1)$ Kac-Moody symmetry of our Hamiltonian Eq. (\ref{ham-4}), which contains $Z_3$ subgroup, can help determine the t'Hooft anomaly.
Importantly, we observe that the possibility of having a $Z_3$ anomaly puts bounds on central charge $c$ and dimension $\Delta$. Using the conformal bootstrap in Refs.~\cite{Lin-2019, Lin-2021} the region in the ($c, \Delta$) plane where the models are $Z_3$ anomalous was identified. It is straightforward to see that our parameters fulfill that bound. Hence, the corresponding CFT agrees with the results of Refs.~\cite{Lin-2019, Lin-2021}.

Generally speaking, one could expect various CFT candidates for our edge theory. One approach could be based on the assumption that since the three-state Potts model with $Z_3$ symmetry at the critical point has $\mathcal{W}_3$ conformal invariance 
\cite{Zamolodchikov-W3,Zamolodchikov-1987},  
our edge model might also belong to the same 
class. Then it would be straightforward to identify the CFT with our central charge from the list of central charges
of minimal $\mathcal{W}_3$ conformal field theories, which can be defined as cosets of the type
$\frac{SU_k(3)\times SU_1(3) }{SU_{k+1}(3)}$.
However $Z_3$ in $\mathcal{W}_3$ CFT forms a center, hence, according to \cite{Chang-2019}, it can not be t'Hooft anomalous
unless it is formed as a mixed anomaly with outer automorphism  \cite{Alavirad-2021, Zhou-2019}. Even though this is an unlikely scenario, we cannot exclude this possibility, which needs further investigation.
 

One can also look for a CFT realizing larger $Z_3 \times Z_3$ symmetry of the Hamiltonian Eq.~(\ref{ham-4}). Then the conformal theory of symmetry $\frac{SU_3(2)}{U(1)}\times \frac{SU_3(2)}{U(1)}$, which is a direct combination of two critical three-state Pots models \cite{FZ-1991} could be na{\"i}vely considered as a CFT candidate, containing a primary field of dimension $2/3$, as we have. 
This scenario emerges upon applying the mean-field approximation to our boundary Hamiltonian. Here, one replaces the operators $X^\pm_p$ in a three-point interaction term in (\ref{ham-4}) with their vacuum averaged values $\langle V|X^\pm_p | V \rangle= \text{const}$. This approximate treatment yields a set of two three-state Potts models defined on odd and even lattice points, respectively. This situation is similar to the edge mode structure in the $Z_2$ Levin-Gu model, where the boundary XX model can be treated as two $Z_2$ Ising models. However, the central charge, which  we have observed numerically $ c\simeq 1.73 $  is far from $ c=1.6 $ CFT of two critical three-state Pots models,  indicating that this is also not our CFT.
We note that as in the $Z_2$ Levin-Gu model \cite{Levin-Gu}, the hidden $U(1)$ symmetry will likely play crucial role here.

The other candidate one could have expected is the $SU(3)_1$ WZNW model. In Ref. \cite{Lanzetta-2022} it was argued, that $SU(3)_1$ low energy theory can be responsible for 
$Z_3 \times Z_3 \times Z_3$ 'tHooft anomaly. 
Even though there could be a matching between the scaling dimension Eq. (\ref{numerics}) with the lightest primary field of the $SU(3)_1$ WZNW model, the latter has central charge $c=2$~\cite{Philippe}, which is considerably larger from what our calculations show. 

\subsection{Conjectured CFT: the coset $SU(3)_2/SU(2)_2$ model}

Here, we look for and identify the low energy CFT of our edge Hamiltonian within the coset models with anomalous $U(1)$ subgroup (\ref{Hw}) and central charge $\sim 1.73$. To this end, we consider the $SU(3)_k/SU(2)_k$ coset CFT. The levels $k$ of Kac-Moody currents in the nominator and denominator here should be the same for denominator algebra to be a subalgebra of the nominator one. Also, $SU(3)_k$ in the nominator contains one more Cartan $U(1)$ subgroup, which is anomalous and can be identified with our $U(1)$ symmetry. 
Furthermore, it is straightforward to calculate the central charge
\bea
\label{cc}
c_k= c_{SU(3)_k}-c_{SU(2)_k}=\frac{8 k}{k+3}-\frac{3 k}{k+2},
\ena
yielding $c_2=1.7$ at level $k=2$, which agrees with our numerical value of $c\simeq 1.73$ within a high precision of $O(10)^{-2}$. 

Because this anomalous U(1) symmetry contains $Z_3$ as a subgroup, one can identify it with $Z_3$ 't Hooft anomaly. Of course, this is our expectation, and it should be verified, for example, by analyzing the modular properties of the partition function of our edge model or by other methods \cite{Nayak-2014,Levin-2021}. In the case of the positive outcome, the partition function will be defined by the Wess-Zumino-Witten type action for $SU(3)_k/SU(2)_k$ coset CFT reflecting the anomaly. Here, we take the viewpoint and conjecture that this coset CFT is the low energy CFT theory for our edge modes and its study will be one of our next goals. 

In the end, we also would like to emphasize that unambiguously identifying CFT with the low energy modes of our Hamiltonian (for example, computing the central charge with very high precision) requires additional precise numerical probes and a better computing infrastructure.

\subsection{Further perspectives for the phase transitions in the edge Hamiltonian}
One can make another duality transformation, similar to the one made in the $Z_2$ Ising model,
which will simplify the form of the edge Hamiltonian (\ref{ham-4}). Namely, let us define
\bea
\label{dual-3}
X^+_p= Y_p^+ Y_{p+1}^-,\qquad Z^+_p=\prod_{k\leq p} W_k^+.
\ena
Then, the edge Hamiltonian acquires the nearest-neighbor coupling form
\bea
\label{ham-5}
\hspace{-0.5 cm}H^{(b)}_{edge}=-\sum_p \big(Y_p^+ Y_{p+1}^-+H.c.\big)\big(1+W^+_{p}W^+_{p+1}+W^-_{p}W^-_{p+1}\big)
\ena

One can note here that this Hamiltonian (and the previous versions of the Hamiltonian describing the same boundary model of our SPT system) does not belong to the integrable set of Fateev-Zamolodchikov models
\cite{FZ-1980, FZ-1991}, nor chiral Potts model \cite{Bazhanov-1990, Chiral-Potts}. 

Now, we would like to generalize some of the points that have been investigated in the numerical simulations of SPT phases in one or two-dimensions
\cite{Yoshida-2015}, \cite{Yoshida-2016-1}, \cite{Yoshida-2017}, \cite{DHLee-2017} and to propose scenarios connected with the observed SPT phase. The symmetric $1D$ edge Hamiltonian operators $H_{edge}$, corresponding to the open boundary of a 2D SPT model, acquire critical transitions. These $H_{edge}$ operators can be described by the sums of the Hamiltonian operators, which correspond to different nontrivial 1D SPT phases with respect to the symmetry transformations induced from the symmetry of the $2D$ model and surviving at the edge. An example is the model corresponding to $H^b_{edge}$.  Let us recall the  $(1+1)$-model with $Z_3\times Z_3$ symmetry (see e.g. Refs.~\cite{Motrunich-2014,DHLee-2017}),
which has three nontrivial SPT phases ($H^2(Z_3\times Z_3, U(1))=Z_3$), corresponds to the following Hamiltonians:
\bea
\label{z3xz3}
H_0&=&-\sum_i\left(X^+_{2i}+X^+_{2i+1}+H.c.\right),
\\
H_1&=&-\sum_i\left(Z^-_{2i-1}X^+_{2i}Z^{+}_{2i+1}+Z^{+}_{2i}X^+_{2i+1}Z^-_{2i+2}+H.c.\right),
\\
H_2&=&-\sum_i\left(Z^{+}_{2i-1}X^+_{2i}Z^-_{2i+1}+Z^-_{2i}X^+_{2i+1}Z^{+}_{2i+2}+H.c.\right).
\ena
Let us note that the following interpolating Hamiltonian (in the notations of \cite{DHLee-2017}) between the presented different phases,
\bea H(\lambda_0,\lambda_1,\lambda_2)=\lambda_0 H_0+\lambda_1 H_1+\lambda_2 H_2,\ena
coincides with the edge Hamiltonian $H^{b}_{edge}$ at the critical point $\lambda_0=\lambda_1=\lambda_2$. In Ref.~\cite{DHLee-2017},  an analytical derivation is presented showing that the interpolating Hamiltonian between $0$ and $1$-phases at the critical point ($\lambda_0=\lambda_1=\frac{1}{2},\;\lambda_2=0$) can be described by two decoupled critical  $Z_3$ clock Potts models, and must be described by the conformal charge $c=2 \times \frac{4}{5}$. The similar derivations could be also done for the interpolating situations  ($\lambda_0=\lambda_2=\frac{1}{2},\;\lambda_1=0$) and ($\lambda_1=\lambda_2=\frac{1}{2},\;\lambda_0=0$).
For the general case of $Z_N$, there are $N$ different SPT phases, and the transitions between the phases corresponding to the "non-adjacent" topological classes will  be described by the successive transitions between the phases corresponding to the adjacent classes.

\subsection{On the $Z_3$ SPT paramagnet}

In this section, we discuss the generalization of the construction of the $Z_2$ SPT Ising paramagnet \cite{Levin-Gu} to the $Z_3$ case. We  present the unitary transformation of the $Z_3$ Potts model defined by the nontrivial element
of the cohomology group $H^3(Z_3,U(1))=Z_3$.  We have already presented above the unitary operator
\begin{equation}
	\label{UUU}
	U=\prod_{\Delta } \nu_3(0,n_1^\Delta,n_2^\Delta,n_3^\Delta)^{S(\Delta)}=\prod_{\Delta } \omega_3(n_1^\Delta,n_2^\Delta-n_1^\Delta,n_3^\Delta-n_2^\Delta)^{S(\Delta)},
\end{equation}
with $\omega_3(n_1,n_2,n_3)=n_1 n_2 n_3 (n_1+n_2)$, which produces
\begin{equation}
	\label{UUUU}
	U= \prod_{\Delta} \varepsilon^{S(\Delta) n_1n_2n_3(n_2-n_1)}.
\end{equation}
One can check that $U$ is symmetric under Ising transformation, $X^+ U X^-=U$, but it can not be reduced to the product 
$U=\prod U_0$ of local and Ising symmetric matrices, $U_0$.

The operator $U$ given by Eq.~(\ref{UUUU}) is one of the elements of the cohomology group $H^3(Z_3, U(1))=Z_3$. Other elements are presented in (\ref{UU}), and they can be obtained by permutations $S_{12}$ and $S_{23}$. These elements are symmetric under Ising transformation $X$ defined by (\ref{Ising-1}).
However, formulating the boundary modes defined by the SPT states (\ref{UUU}) is a task for future study.

\subsection{Concluding remarks}
We have constructed the $Z_3 \times Z_3 \times Z_3$ and $Z_3$ SPT models in two spatial dimensions with gapless edge modes based on the three-state Potts model in the paramagnetic phase. To derive the SPT model, we extended the construction on the $Z_2$ Ising model by Levin and Gu to the case of $Z_3 \times Z_3 \times Z_3$ and $Z_3$ symmetries, even though the same result could have been achieved upon employing the general method put forward in Ref.~\cite{Wen-2013}. Here, we find various edge modes based on a larger possibility of identifying external boundary spin states. Since the external spins in our construction may have various configurations,
we realize that they will form different edge models whose investigation opens an area for future investigations.  

We show that the obtained boundary models have the property of self-triality, indicating the gapless nature of the modes. The SPT phases we derived, according to the classification of Ref.~\cite{Wen-2013},
are defined by $H^3(Z_3, U(1))=Z_3$ and $H^3(Z_3 \times Z_3 \times Z_3, U(1))=(\times Z_3)^7$, and the cohomology groups of $Z_3$ and $Z_3 \times Z_3 \times Z_3$ symmetry groups with $U(1)$ coefficients respectively. 
The corresponding gauged phases carry anionic properties and illustrate non-standard statistics.
Our numerical DMRG calculations show that the  boundary model has a central charge $c\simeq1.73$ and smallest conformal dimension  $\Delta = 2/3$. It also has a self-triality property, directly indicating the model is gapless.  Moreover, the boundary model appears to have a larger symmetry, namely $S_3$. The group $S_3$ also permutes SPT states belonging to different blocks of the cohomology group $H^3(Z_3 \times Z_3 \times Z_3, U(1))$. 

We also have revealed a "hidden" symmetry of the edge model of the derived $Z_3 \times Z_3 \times Z_3$ SPT paramagnet, generated with a topological winding number. This observation implies that the spectrum of the 1D edge model should be described not only by energy and momentum but also by the characteristic winding number. Importantly, this hidden $U(1)$ symmetry fulfills holomorphic factorization condition and generates Kac-Moody
algebra, which is anomalous and may carry t'Hooft anomaly. Based on this result, we have analyzed known conformal field theories, which potentially may define low energy limit of our edge Hamiltonian. Our candidate for low-energy CFT is the coset $SU(3)_2/SU(2)_2$, which fits well with all our numerical simulations.

\section*{Acknowldgements}
We are grateful to A. Sedrakyan, H. Babujian, A. Belavin, and A. Litvinov for helpful discussions. We also acknowledge communications from Ryan Lanzetta and Juven Wang with thanks. 
The research was supported by startup funds from the University of Massachusetts, Amherst (TAS) and
Armenian SCS grants Nos.   20TTAT-QTa009 (HT, SK, TH), 20TTWS-1C035 (SK, TH), 21AG-1C024 (HT, MM, SK), and
21AG-1C047 (TH). We are grateful to the anonymous reviewer of JHEP for their insightful comments.

\begin{appendix}

	\section{Permutation group $S_3$}
 
	The permutation group has six generators, which are
	\bea
	\label{S3}
	X^+&=&
	\left(
	\begin{array}{ccc}
		0&1&0\\
		0&0&1\\
		1&0&0
	\end{array}
	\right),
	\quad
	X^-= (X^+)^\dagger=
	\left(
	\begin{array}{ccc}
		0&0&1\\
		1&0&0\\
		0&1&0
	\end{array}
	\right),
	\nn \\
	S_{12}&=&
	\left(
	\begin{array}{ccc}
		0&1&0\\
		1&0&0\\
		0&0&1
	\end{array}
	\right),\nn
	\quad
	S_{23}=
	\left(
	\begin{array}{ccc}
		1&0&0\\
		0&0&1\\
		0&1&0
	\end{array}
	\right),
	\\
	S_{13}&=&
	\left(
	\begin{array}{ccc}
		0&0&1\\
		0&1&0\\
		1&0&0
	\end{array}
	\right),
	\quad
	I=
	\left(
	\begin{array}{ccc}
		1&0&0\\
		0&1&0\\
		0&0&1
	\end{array}
	\right).
	\ena
	Any two elements of the group, besides the identity $I$, generate the whole group.
	One can see, for example, that various products of $S_{12}$ and $S_{23}$
	produces all elements in $S_3$.
	\begin{align}
		\label{A1}
		&S_{12}S_{23}=X^-,
		\quad S_{23}S_{12}=X^+,
		\quad S_{23}S_{12}S_{23}=S_{13},
		\nn
		\\
		&(X^\pm)^2 = X^\mp,\quad  (X^\pm)^3=S_{ij}^2=I.
	\end{align}
	Straightforward calculations show
	\bea
	\label{A2}
	S^{-1}_{12}X^\pm S_{12}= X^\mp,
	\\
	S^{-1}_{23}X^\pm S_{23}= X^\mp
	\nn,
	\ena
	which indicates, that paramagnetic Hamiltonian (\ref{ham}) commutes with the whole group
	$S_3$.
	Moreover, taking into account
	\bea
	\label{A3}
	S^{-1}_{12}Z^\pm S_{12}&=&\varepsilon^{\pm} Z^\mp \nn\\
	S^{-1}_{23}X^\pm S_{23}&=&\varepsilon^{\mp} X^\mp\nn\\
	S^{-1}_{13}Z^\pm S_{13}&=& Z^\mp,
	\ena
	one realizes that both edge Hamiltonian operators $H_{edge}^{(a/b)}$ have $S_3$ symmetry.

	\section{Nontriviality of 3-cocycles}
	\label{sec:nontriviality}
	
	To check that 3-cocycles \eqref{3-cocycle-1}, \eqref{3-cocycle-2} and \eqref{3-cocycle-3} are not trivial, we need to solve an equation $\omega_3 = \delta \omega_2$ for $\omega_2$. One can straightforwardly do that using Wolfram Mathematica. Take the basis in the space $C^2$ of 2-forms consisting of all possible monomials.
	$$
	(n_1^{(1)})^{p_1^{(1)}}(n_1^{(2)})^{p_1^{(2)}}(n_1^{(3)})^{p_1^{(3)}}(n_2^{(1)})^{p_2^{(1)}}(n_2^{(2)})^{p_2^{(2)}}(n_2^{(3)})^{p_2^{(3)}}
	$$
	with the powers
	$p_i^{(j)}=0,1,2$.
	There are $3^6=729$ such monomials. Applying the coboundary operator $\delta$ to each, we obtain a set $E$ of $3$-forms spanning all the exact 3-cocycles. What is left to do is check if a given $3$-form can be expressed as a linear combination of them. One way to approach this is to decompose all of the polynomials in $E$ into vectors consisting of coefficients before each of the monomials. This gives us a matrix $A$, each column of which represents one of the polynomials in $E$. After decomposing $\omega_3$ into vector $P$ in a similar manner, we get a system of linear equations,
		$AX=P$,
	which is solved using standard tools.

\end{appendix}





\end{document}